\documentclass[twocolumn,aps,showpacs,amsmath,superscriptaddress]{revtex4-2}
\usepackage[latin9]{luainputenc}
\usepackage{mathtools}
\usepackage{amsmath}
\usepackage{amssymb}
\usepackage{graphicx}
\usepackage[symbol]{footmisc}

\makeatletter

\usepackage{dcolumn}
\usepackage{bm}
\usepackage{ifpdf}
\usepackage{lineno}\usepackage{amsfonts}
\usepackage{bbm}
\usepackage{mathrsfs}
\usepackage{upgreek}
\usepackage{epstopdf}
\usepackage{setspace}
\usepackage[matrix,frame,arrow]{xypic}
\usepackage{rotating}
\usepackage{float}
\usepackage{hyperref}
\usepackage{color}
\definecolor{med-blue}{RGB}{25,25,112}
\hypersetup{colorlinks, linkcolor={blue},citecolor={blue}, urlcolor={blue}}

\newcommand{\ket}[1]{\vert{#1}\rangle}

\makeatother

\begin{document}
	
	\title{Imaging of microwave magnetic field orientation using continuous-wave experiments on nitrogen-vacancy centers in diamond}
	\author{Akshat Rana}
	\thanks{These authors contributed equally to this work.}
	\affiliation{Department of Physics, Bennett University, Greater Noida 201310, India}
	\author{Pooja Lamba}
	\thanks{These authors contributed equally to this work.}
	\affiliation{Department of Physics, Bennett University, Greater Noida 201310, India}
	\author{Atanu Ghosh}
	\affiliation{Department of Physics, IIT Madras, Chennai, India}
	\author{Siddharth Dhomkar}
	\affiliation{Department of Physics, IIT Madras, Chennai, India}
	\affiliation{Center For Quantum Information, Communication And Computing, IIT Madras, Chennai, India}
	\author{Rama K. Kamineni}
	\email{koti.kamineni@gmail.com}
	\affiliation{SIAS, Krea University, Sri City 517646, India}

	\date{\today}
	\begin{abstract}
		{Imaging of microwave magnetic fields with nano-scale resolution has interesting applications. Specifically, detecting the orientation of the microwave fields is useful in condensed matter physics and quantum control. However, most of the existing methods for microwave field imaging are limited to detecting the magnitude of the fields. Due to their small sensor size and favorable optical and spin properties, nitrogen-vacancy (NV) centers in diamond are highly suitable for imaging dc and ac magnetic fields. The reported methods for detecting the orientation of microwave magnetic fields use pulsed Rabi frequency measurements. Here, we demonstrate imaging of the orientation of microwave magnetic fields by only using continuous-wave experiments on NV centers. This simplifies the sensor apparatus and is particularly advantageous in applications where pulsing of the target microwave field is not possible. The method requires static bias magnetic field oriented perpendicular to the quantization axis of NV centers. 
		We detect the direction of an arbitrary microwave magnetic field using NV centers of two different orientations. Moreover, we demonstrate that the projection of the microwave fields onto a plane can be imaged using NV centers of single orientation. It can be straightforwardly implemented using a single NV center.
			
		} 
	\end{abstract}
	
	
	\maketitle

\section{Introduction}
Microwave devices have become integral parts of new age quantum technologies such as quantum information processing and quantum sensing. Microwave electric and magnetic fields are used to manipulate quantum systems like superconducting qubits \cite{Wallraff2004SCQ,You2011SCQrev}, spins in solid-state \cite{Jelezko2004eRabi} and quantum dots \cite{Vandersypen2006Qdot}. Knowledge of the distribution of the control fields at the site of quantum systems is useful in extracting optimal performance from them. However, this requires precise detection of the microwave fields at micro- and nano-scales. 
Microwave field imaging is also useful in condensed matter physics for studying magnetic excitations such as spin waves \cite{Yacoby_RevNVCondMat2018}. Various techniques based on different architectures have been developed for precise microwave fields detection \cite{Clarke_RFSQUID1995,vanderWeide_RFtip1997,Lee_MWmicroscope2000,Boehi_MWimgAtoms2010,Boehi_MWimgVapor2012,Ockeloen_MW_BECmag2013,vanderSar_SpinWaves2015,Maletinsky_MWsens2015njp,Koti2020_LAC}. Among these architectures, nitrogen-vacancy (NV) centers in diamond have the advantages of small sensor size and ability to operate under ambient conditions.

NV centers are point defects in diamond with ${\textrm{C}_{3V}}$ symmetry \cite{RevDoherty,SuterNVRev2017}. The electron and nuclear spins of these centers are employed in a number of applications ranging from quantum information processing to nano-scale imaging of biological molecules \cite{RevDegen,RevJacques,RevWalsworth,RevChildress,WrachtrupESens}. NV centers are very good sensors of both dc and ac magnetic fields \cite{JW2008Nat,Degen_Mag2008APL,Lukin_Mag2008Nat,Staudacher2013,DegenScience2017,JelezkoScience2017}. Moreover, vector detection schemes, where both strength and orientation of the applied field is determined, are thoroughly investigated for dc magnetic field sensing \cite{Awschalom_VecMag2010,Fang_VecMag2018,Clevenson_VecMag2018,Walsworth_VecMag2018,Budker_VecMag2020,Hollenberg_VecMag2020,Weggler_VecMag2020,Du_VecMag2020,Flatae_VecMag,Yacoby_AngleSens2021npj}. However, in the microwave frequency regime, most of the sensing protocols based on NV centers and other architectures lack vector detection capabilities. Detecting the orientation of the microwave field is important for many of the applications, e.g., optimal control of $^{13}$C nuclear spin qubits coupled to an NV center, where the bias magnetic field is aligned non-parallel to the NV axis, requires knowledge of the orientation of the control field.

Conventionally, vector detection of microwave fields can be achieved by resonantly driving the transitions of three NV centers of different orientations \cite{DuVectorMW}. Vector detection of radio-frequency fields by a single NV center has been achieved in Refs. \cite{Cappellaro_VecAC2021,Jingfu_RFsens2023}, however, extending them to microwave regime is difficult. Recently, vector detection of ac fields by a single NV center has been demonstrated by applying the bias field perpendicular to the NV axis \cite{PoojaVecSens2024}. A similar method that uses resonant Rabi oscillations of a single NV center under strong local strain fields and zero bias field has also been demonstrated \cite{WangVecMWmag2024}.
All these methods are based on pulsed experimental techniques. 
While the pulsed schemes offer highest sensitivities, the most straightforward approach to magnetometry relies on the continuous-wave optical detection of magnetic resonance (CW-ODMR) technique.
In this paper, we focus on a CW-ODMR method for imaging the orientation of microwave magnetic fields. It is based on the theory developed in Ref. \cite{PoojaVecSens2024}. 
When the bias magnetic field is applied perpendicular to an NV axis, the dipole moments of the transitions between the energy levels of the electron spin are oriented perpendicular to each other. By resonantly exciting these transitions in CW-ODMR experiments and using their intensities, orientation of the microwave fields can be measured. Two NV centers of different orientations are needed to image arbitrary orientations of microwave fields in three dimensions. However, the projections of microwave fields onto a plane can be imaged by a single NV center. The experiments have been performed by using ensemble NV centers. Though there are NV centers of four different orientations in an ensemble, only transitions of specific orientations are addressed in the ODMR experiments.

This paper is organized as follows. We describe the method in Sec. \ref{Method}, and its experimental implementation in Sec. \ref{Expt}. Finally, in Sec. \ref{Conc}, we summarize the key outcomes.

\section{Methodology}
\label{Method}
\begin{figure*}
	\includegraphics[width=16cm]{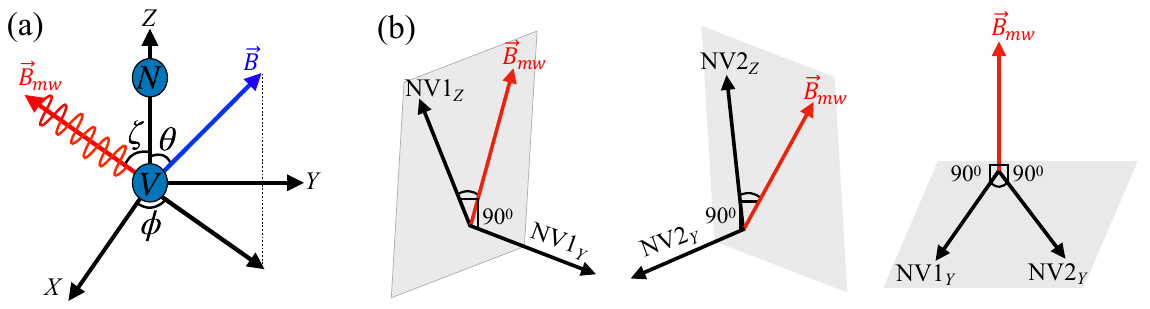} \caption{(a) Coordinate system fixed to an NV center along with the static ($\vec{B}$) and microwave ($\vec{B}_{mw}$) magnetic field orientations. (b) Axes of two NV centers of different orientations and the microwave vector. NV1$_Z$ and NV2$_Z$ represent the $Z$-axes of the centers and NV1$_Y$ and NV2$_Y$ represent the $Y$-axes.}
	\label{Axes} 
\end{figure*}
NV center in diamond has spin-1 ($S=1$) electronic ground and excited states and singlet intermediate states. The electronic spin of the center can be polarized into the $m_S=0$ state via optical pumping and the spin state dependent fluorescence emitted by the center is used to read its spin state \cite{SuterNVRev2017}. There are four possible orientations for an NV center in diamond with respect to the crystal structure. As shown in Fig. \ref{Axes}(a), we define an NV coordinate system fixed to one of the four NV orientations.
$Z$-axis is defined to be along the NV axis. The microwave magnetic field $\vec{B}_{mw}$, whose orientation we need to determine, makes an angle $\zeta$ with the $Z$-axis. The projection of the microwave field vector into the transverse plane is defined as $X$-axis. $Y$-axis is chosen perpendicular to both $X$- and $Z$-axes. $\theta$ and $\phi$ represent the polar and azimuthal angles of the static magnetic field $\vec{B}$ in this coordinate system.
The Hamiltonian of the ground state of the electron spin of an NV center under applied static magnetic field can be written as 
\begin{align}
	{\cal H}_{0}= & DS_{z}^{2}+\gamma_{e}B(\sin\theta \cos\phi \ S_{x}+ \sin\theta \sin\phi \ S_{y}+\cos\theta\ S_{z}),\label{eq:Hamiltonian}
\end{align}
where $S_{x/y/z}$ represent the components of the spin-1 angular momentum operator. We set $h=1$ and use frequency units for energy. $D=2870$ MHz is the zero-field splitting of the electron spin. $\gamma_{e}$ and $B$ represent the gyromagnetic ratio of the electron spin and the strength of the applied static magnetic field respectively.

The Hamiltonian corresponding to the interaction of the applied microwave magnetic field with the electron spin of an NV center can be written as
\begin{align}
	{\cal H}_{mw}= & \gamma_{e} B_{mw}(\sin\zeta\ S_x+\cos\zeta\ S_z) \cos(\omega t+\varphi),
\end{align}
where $B_{mw}$ represents the amplitude of the microwave field. $\omega$ and $\varphi$ represent the frequency and phase angle of the field.

When $\theta=0$, i.e., the static field is aligned with the NV axis, the eigenstates of ${\cal H}_0$ are $\ket{m_S}=\ket{0}$, $\ket{-1}$, and $\ket{1}$, where $m_S=-1$, $0$, and $1$ are the eigenvalues of the operator $S_z$.
The transitions between the states $\ket{0}$ and $\ket{-1}$, and  $\ket{0}$ and $\ket{1}$ are allowed while the transition between the states $\ket{-1}$ and $\ket{1}$ is forbidden.
When $\theta=\pi/2$ and for field strengths of the order of 10 mT or less, the eigenstates of ${\cal H}_0$ can be approximately written as $\ket{0}$, $\ket{-}$, and $\ket{+}$, where $\ket{-}=\frac{\ket{-1}-\ket{1}}{\sqrt{2}}$,
and $\ket{+}=\frac{\ket{-1}+\ket{1}}{\sqrt{2}}$ \cite{Genovese_TempSens}. All the three transitions between the states $\ket{0}$, $\ket{-}$, and $\ket{+}$ are allowed and their dipole moments are oriented perpendicular to each other.
In Ref. \cite{PoojaVecSens2024}, the Rabi oscillations of these transitions have been utilized to determine the strength and orientation of an unknown ac field. However, measurement of Rabi oscillations requires pulsing of the ac field and it is not always possible to pulse the ac field that is to be detected.

Here, we use the electron spin transitions at $\theta=\pi/2$ to image the orientation of a microwave magnetic field by only performing CW-ODMR measurements. 
In zero static magnetic field, the transition frequencies of $\ket{0}\longleftrightarrow\ket{-}$ and $\ket{0}\longleftrightarrow\ket{+}$ transitions are around 2.87 GHz and they increase non-linearly with the transverse field strength.
A CW-ODMR spectrum can be recorded by sweeping the frequency of the microwave field for a fixed static field or by sweeping the strength of the static field for a fixed microwave field, while continuously monitoring the intensity of the fluorescence emitted by NV centers.
A dip in the fluorescence intensity occurs when the microwave frequency matches with one of the transition frequencies \cite{Bitter_ODMR1949,Suter_ODMRRev2020}.
The amplitude of the dip in the intensity depends on the strength of the microwave field and on the dipole moment of the corresponding transition.
As given in Ref. \cite{PoojaVecSens2024}, when the static field $\vec{B}$ is oriented transverse to the NV-axis, the dipole moment of the transition $\ket{0}\longleftrightarrow\ket{+}$ ($\ket{0}\longleftrightarrow\ket{-}$) is maximum (minimum) in the direction of $\vec{B}$ and it is minimum (maximum) in the direction perpendicular to both $\vec{B}$ and NV-axis. Correspondingly, the intensity of the transition $\ket{0}\longleftrightarrow\ket{+}$ in the ODMR spectrum is maximum when $\vec{B}$ is oriented along $X$-axis and it is minimum when $\vec{B}$ is oriented along $Y$-axis.
This implies that the $Y$-axis, which is perpendicular to the plane containing the NV axis and the microwave magnetic field vector, can be determined by finding the static field orientation in the plane perpendicular to the NV axis where the transition $\ket{0}\longleftrightarrow\ket{+}$ has minimum intensity.


Now, consider two NV centers having different orientations, labeled as NV1 and NV2, and illustrated in Fig. \ref{Axes}(b). Assume that the microwave field $\vec{B}_{mw}$ interacts with both these centers. 
The axes corresponding to the coordinate systems of these two centers are labeled as NV1$_{X/Y/Z}$ and NV2$_{X/Y/Z}$. NV1$_Z$ and NV2$_Z$ represent the directions of the corresponding NV axes. As shown in Fig. \ref{Axes}(b), NV1$_Y$ is perpendicular to the plane containing NV1$_Z$ and $\vec{B}_{mw}$, and NV2$_Y$ is perpendicular to the plane containing NV2$_Z$ and $\vec{B}_{mw}$. This implies that the direction of $\vec{B}_{mw}$ is perpendicular to the plane containing NV1$_Y$ and NV2$_Y$. The directions NV1$_Y$ and NV2$_Y$ can be determined from the static field orientations where the transition $\ket{0}\longleftrightarrow\ket{+}$ has minimum intensity for each of the two centers as described earlier in this section. The cross product of NV1$_Y$ and NV2$_Y$ directions provides the direction of the microwave magnetic field $\vec{B}_{mw}$. Hence, two NV centers of different orientations are sufficient to determine the orientation of an arbitrary microwave field. 
The projection of the microwave field onto a plane or the direction of a microwave field, if it is known to be confined to a plane, can be detected by using only one NV center.
Since the minimum frequency of the electron spin transitions used here is 2.87 GHz, this method is not applicable for frequencies below 2.87 GHz.

\section{Experiments}
\label{Expt}

All the experiments are conducted on a home-built confocal microscope setup equipped with microwave electronics. A CVD grown diamond crystal with a nitrogen concentration of 900 ppb is used. A diode pumped solid state laser of wavelength 532 nm is used for excitation of NV centers. The fluorescence emitted by the NV centers is separated from the excitation path by a 562 nm dichroic mirror. It is further filtered by using a 650 nm long-pass filter and focused onto a single-photon detector.
The experiments require precise orientation of static magnetic field with respect to the diamond crystal. 
For this, we use two rotation stages and position them in such a way that their axes cross at the site of the diamond sample \cite{ShimArXiv}. By attaching a permanent magnet to these stages and rotating it, a three dimensional rotation of the magnetic field with respect to the diamond crystal can be achieved.
The strength of the static field at the site of the sensor is 10.2 mT.
The microwave fields, whose orientation we are trying to determine, are applied by sending microwave current through a copper wire of 25 $\mu$m diameter attached to the diamond surface.

\begin{figure}
	\includegraphics[width=8.6cm]{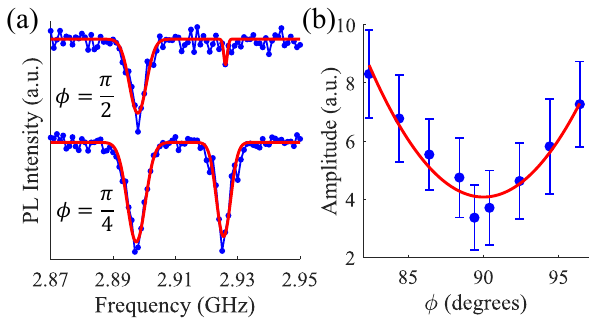} 
	\caption{ (a) CW-ODMR spectra for transverse static field ($\theta=\pi/2$), and $\phi=\pi/2$ and $\pi/4$. (b) Intensity of the $\ket{0}\longleftrightarrow\ket{+}$ transition (right peak in (a)) as a function of $\phi$. Data is fit to the function $a \cos^2(\phi)+b$. The minimum of the fit corresponds to $\phi=\pi/2$. Error bars represent 95\% confidence intervals.}
	\label{ODMR}  
\end{figure}

From the nitrogen concentration of the diamond sample and the intensity of the emitted fluorescence, we can conclude that there are hundreds of NV centers in the excitation volume of the confocal setup. The fluorescence emitted by NV centers of all the four orientations contributes to the total detected fluorescence. However, with appropriate orientation of the static magnetic field, the electron spin transitions of NV centers of different orientations can be resolved in frequency space. Hence, the spin transitions of each NV orientation can be addressed separately. 
CW-ODMR spectra of electron spin transitions are recorded by sweeping the frequency of the microwave fields while continuously measuring the intensity of the fluorescence signal. 

When a static field of strength 10.2 mT is applied perpendicular to an NV axis, the transition frequencies of the $\ket{0}\longleftrightarrow\ket{-}$ and $\ket{0}\longleftrightarrow\ket{+}$ transitions are 2898 and 2926 MHz respectively. The intensities of these transitions in the ODMR spectra depend on the angle ($\phi$) between the transverse component of the microwave field and the direction of the static field in the transverse plane. As discussed in Sec. \ref{Method}, the transition $\ket{0}\longleftrightarrow\ket{+}$ has minimum intensity when the static field is oriented perpendicular to both NV axis and the microwave field.
Fig. \ref{ODMR}(a) shows the ODMR spectra for $\phi=\pi/4$ and $\phi=\pi/2$. At $\phi=\pi/4$, both transitions have almost equal intensity and when $\phi=\pi/2$, the transition $\ket{0}\longleftrightarrow\ket{+}$ has negligible intensity.
The intensity of the transition $\ket{0}\longleftrightarrow\ket{+}$ as a function of $\phi$ is plotted in Fig. \ref{ODMR}(b).
The minimum of the curve fit to the experimental data corresponds to $\phi=\pi/2$.
The corresponding direction of the static field is perpendicular to both NV axis and the microwave field. Consistent with Sec. \ref{Method}, we label this direction as NV1$_Y$. Next, the static field is oriented perpendicular to another NV axis and the ODMR measurements described above are repeated to determine the direction NV2$_Y$, which is perpendicular to the second NV axis and the microwave field.
The direction of the microwave field ($\vec{B}_{mw}$) of frequency 2926 MHz is determined from the cross-product of the directions NV1$_Y$ and NV2$_Y$.

\begin{figure}
	\includegraphics[width=8.6cm]{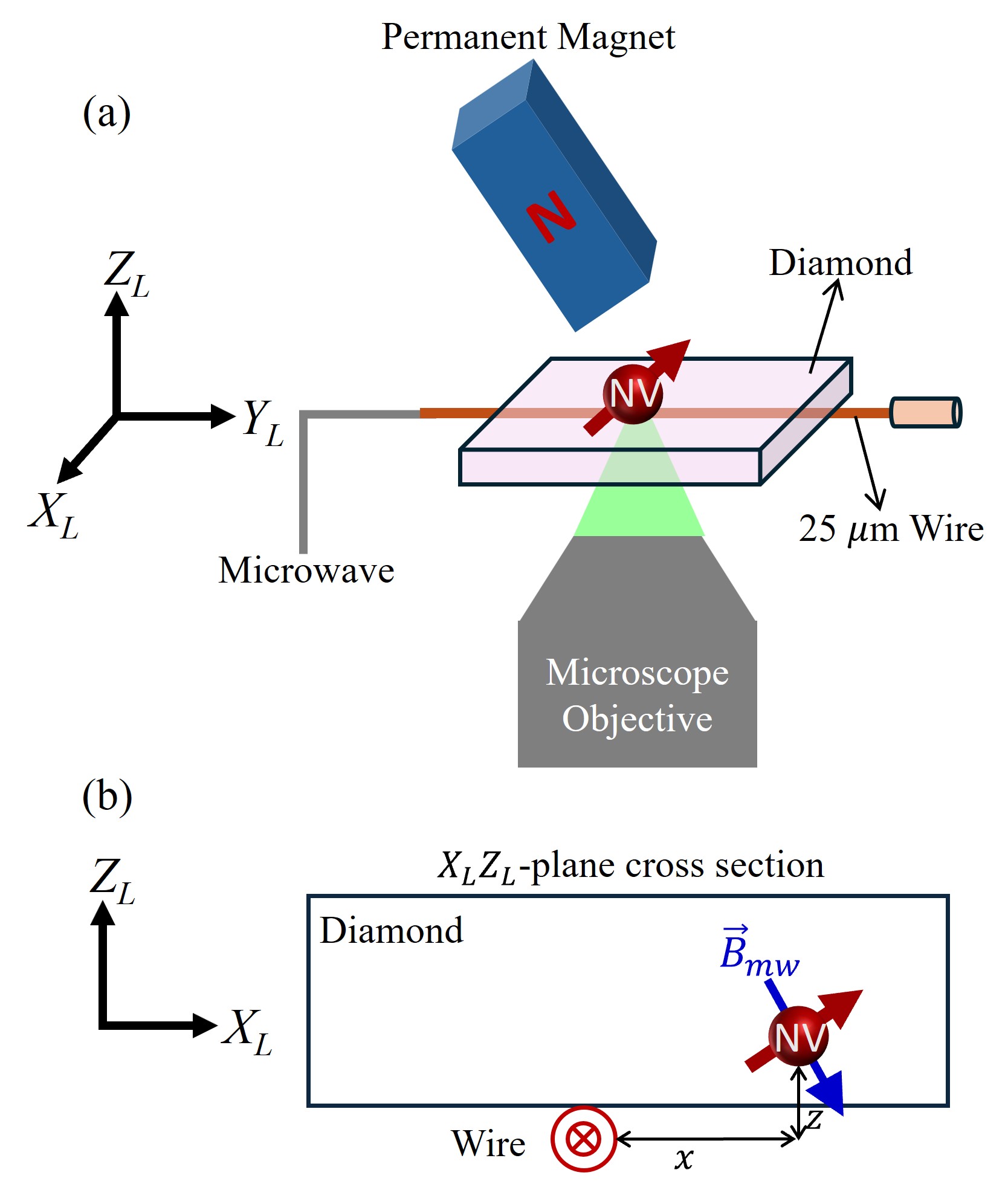} \caption{(a) Schematic of the geometry of the diamond crystal, the wire, and the lab coordinate system. (b) $X_LZ_L$-plane cross section of the diamond crystal and the wire. NV sensor position and the microwave magnetic field vector are also indicated.}
	\label{Setup} 
\end{figure}
 
To compare the experimentally determined microwave vector with the theoretically expected one, we define a lab coordinate system with respect to the diamond crystal as shown in Fig. \ref{Setup}(a). Its z-axis ($Z_L$) is defined to be perpendicular to the plane of the diamond crystal, y-axis ($Y_L$) is parallel to the 25 $\mu$m wire attached to the diamond surface, and x-axis ($X_L$) is perpendicular to both $Z_L$ and $Y_L$. In this coordinate system, the experimentally determined directions NV1$_Y$ and NV2$_Y$ are given as $[-0.86, 0.42, -0.29]^T$ and $[0.85, 0.46, 0.25]^T$ respectively. Their cross product gives us the direction of the microwave vector as $[0.30, -0.04, -0.95]^T$.
Now, we can calculate the theoretically expected direction of the microwave vector from the geometry of the diamond crystal, the wire and the NV sensor position. In the lab coordinate system, the orientations of the two NV axes (NV1$_Z$ and NV2$_Z$) that are used in the experiments can be written as $\frac{1}{\sqrt{3}}[-1, -1, 1]^T$, $\frac{1}{\sqrt{3}}[1, -1, -1]^T$. 
As illustrated in Fig. \ref{Setup}(b), the NV sensor is at a position of $x=61\ \mu$m and $z=18\ \mu$m from the center of the wire. The direction of the microwave magnetic field at the site of the sensor is the same as the direction of the tangent of the circular arc placed at the sensor position and centered at the center of the wire. From the $x$ and $z$ values, the direction of the microwave field can be calculated as $[0.28, 0, -0.96]^T$. 
This closely matches the experimentally determined vector and the angular error between the two vectors is $2.6^\circ$.

So far, we have described determining the direction of an arbitrary microwave magnetic field by using NV centers of two different orientations.
However, in the current experiments, the microwave field is applied by sending the microwave current through the wire antenna positioned along $Y_L$-axis and hence the orientation of the resultant field is confined to a single ($X_L Z_L$) plane. 
In this scenario, NV centers of single orientation is sufficient to determine the orientation of the microwave field. 
Consider that the orientation of the corresponding NV-axis is $\frac{1}{\sqrt{3}}[-1, -1, 1]^T$. In the lab coordinate system, the microwave vector can be written as $[\sin\alpha, 0, \cos\alpha]^T$, where $\alpha$ is the angle between the $Z_L$-axis and the microwave vector.
The unit vector that is perpendicular to both these directions can be written as $\frac{1}{\sqrt{2+\sin2\alpha}}[-\cos\alpha, \cos\alpha+\sin\alpha, \sin\alpha]^T$. This unit vector can be determined experimentally as described earlier in this section and we write it as $[a, b, c]^T$. By equating these two vectors, an expression for $\alpha$ can be derived as $\alpha=\frac{1}{2} \sin^{-1}(\frac{2c^2-1}{a^2+b^2})$. Thus, the angle $\alpha$ can be determined from the values $a$, $b$, and $c$.


\setlength{\tabcolsep}{7pt}
\renewcommand{\arraystretch}{1.5}
\begin{table}[h]
	\begin{center}
		\begin{tabular}{| c | c |  c | c |}
			\hline
			\multicolumn{2}{|c|}{Position $(\mu m)$}&\multicolumn{2}{|c|}{Angle $\alpha$ (degrees)}\\
			\hline
		     $x$ & $z$ & Experimental & Theoretical \\	
			\hline
			\hline	
			 47.7 & 16.5 & 161.3  & 160.9  \\
			\hline	
		    47  & 18.5  & 156.9 & 158.5  \\
			\hline	
	        46.3 & 20.0  & 156.6 & 156.6  \\
			\hline
		     45.5 & 22.0 & 154.8  & 154.2 \\
			\hline
			 44.0 & 25.0 & 152.9  & 150.4  \\
			\hline
			 43.0 & 26.6  & 151.6 & 148.3  \\
			\hline		
			 38.6 & 32.5 & 141.2  & 139.9  \\
			\hline	
			 36.9 & 34.5 & 138.1 & 136.8  \\
			\hline	
			38.5 & 26.7 & 145.7 & 145.3  \\
			\hline	
			
		\end{tabular}
	\end{center}
	\caption{Experimentally measured values of $\alpha$, the angle between $Z_L$-axis of the lab frame and the microwave vector, in comparison with the theoretically calculated ones for different sensor positions in the $X_LZ_L$-plane.}
	\label{table}
\end{table}

\begin{figure}
	\includegraphics[width=8cm]{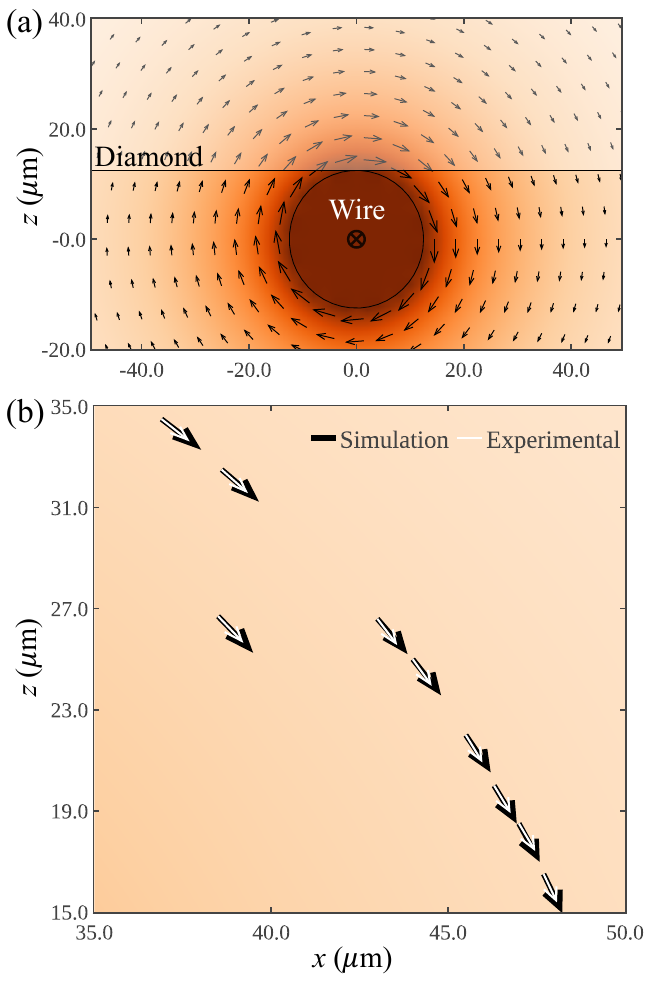} \caption{(a) Simulated microwave field distribution around the wire in the $X_LZ_L$-plane. The microwave current flows into the plane of the paper. (b) Experimentally determined microwave field vectors in comparison with the simulated ones.}
	\label{Result}
\end{figure}

By using one of the four NV orientations of the ensemble NV centers, we have performed imaging of the orientation of the microwave magnetic field generated by the wire antenna in the $X_LZ_L$-plane. The angle $\alpha$ is measured for different values of $x$ and $z$ and the results are tabulated along with the theoretical values in Table \ref{table}.
Simulated orientation of the field is shown in Fig. \ref{Result}(a). As the microwave current flows along $Y_L$-axis, the resultant magnetic field curls around it. The experimentally determined vectors in comparison with the simulated ones are plotted in Fig. \ref{Result}(b). 
The results are in excellent agreement with the simulation. Though these experiments are performed by using ensemble NV centers, they can be straightforwardly implemented by using single NV centers.

Angle sensitivity is an important metric to evaluate the angular resolution achieved by the imaging method.
It can be expressed as $\eta=\frac{\sigma_S \sqrt{nt}}{\frac{dS}{d\phi}}$, where $\sigma_S$ is the uncertainty in the signal $S$, $t$ is the sensing time, and $n$ is the number of repetitions of the experiment. By assuming that the intensities of the transitions in the ODMR spectra are proportional to the square of the corresponding Rabi frequencies, the signal can be written as $S=\tan^2\phi=\frac{I_1}{I_2}$, where $I_1$ and $I_2$ represent the intensities of $\ket{0}\longleftrightarrow\ket{-}$ and $\ket{0}\longleftrightarrow\ket{+}$ transitions respectively. From this, we can write $\sigma_S=\sqrt{2} \tan^2\phi \ \frac{\sigma_{I_1}}{I_1}$, where $\sigma_{I_1}$ is the uncertainty in the measurement of $I_1$. 
After substituting the expressions for $\sigma_S$ and $\frac{dS}{d\phi}$ and simplifying, we get $\eta=\frac{\sin(2\phi)}{2\sqrt{2} }\frac{\sigma_{I_1}}{I_1} \sqrt{nt}$. By taking $\sin(2\phi)=1$, the maximum angle sensitivity can be obtained as $\eta_{\textrm{max}}=\frac{1}{2\sqrt{2} }\frac{\sigma_{I_1}}{I_1} \sqrt{nt}$. From the fitting error of intensities in ODMR $\frac{\sigma_{I_1}}{I_1}=0.08$, and $nt=1$ s, the maximum value of angle sensitivity in our experiments is $28\ \textrm{mrad}/\sqrt{\textrm{Hz}}$. The actual value of sensitivity will be lower than this as the experiments are performed at $\phi=\pi/2$. The sensitivity can be improved by optimizing the ODMR measurements which will reduce the sensing time. For a single NV center close to the diamond surface, assuming a count rate of 200 kcps and a CW-ODMR contrast of 30\%, shot noise limited maximum sensitivity $\eta_{\textrm{max}}$ will be about 2.6 $\textrm{mrad}/\sqrt{\textrm{Hz}}$.

\section{Conclusion}
\label{Conc}
We have imaged the orientation of a microwave magnetic field solely using CW-ODMR measurements. The method offers simplicity and reduced technical requirements compared to the methods based on the pulsed Rabi frequency measurements. It is particularly useful in situations where pulsing of the target magnetic field is infeasible. We have used NV centers of one of the four orientations of an ensemble to image the orientation of a microwave field, whose direction is confined to a plane. 
Importantly, the technique is readily applicable to single NV sensors, making it highly suitable for nano-scale imaging of microwave field orientations using diamond nanopillar probes integrated with an atomic force microscope \cite{Maletinsky_NVAFM2012,Maletinsky_MWsens2015njp}.


\section{Acknowledgements}

A.R. and R.K.K. acknowledge support from Department of Science \&
Technology - Science \& Engineering Research Board (DST-SERB), India
through grant no. SRG/2020/000765. S.D. thanks Indian Institute of
Technology, Madras, India for the seed funding. S.D. acknowledges
the financial support by the Mphasis F1 Foundation given to the Centre
for Quantum Information, Communication, and Computing (CQuICC).

\bibliography{bibNV_LAC}

\begin{thebibliography}{47}%
\makeatletter
\providecommand \@ifxundefined [1]{%
 \@ifx{#1\undefined}
}%
\providecommand \@ifnum [1]{%
 \ifnum #1\expandafter \@firstoftwo
 \else \expandafter \@secondoftwo
 \fi
}%
\providecommand \@ifx [1]{%
 \ifx #1\expandafter \@firstoftwo
 \else \expandafter \@secondoftwo
 \fi
}%
\providecommand \natexlab [1]{#1}%
\providecommand \enquote  [1]{``#1''}%
\providecommand \bibnamefont  [1]{#1}%
\providecommand \bibfnamefont [1]{#1}%
\providecommand \citenamefont [1]{#1}%
\providecommand \href@noop [0]{\@secondoftwo}%
\providecommand \href [0]{\begingroup \@sanitize@url \@href}%
\providecommand \@href[1]{\@@startlink{#1}\@@href}%
\providecommand \@@href[1]{\endgroup#1\@@endlink}%
\providecommand \@sanitize@url [0]{\catcode `\\12\catcode `\$12\catcode
  `\&12\catcode `\#12\catcode `\^12\catcode `\_12\catcode `\%12\relax}%
\providecommand \@@startlink[1]{}%
\providecommand \@@endlink[0]{}%
\providecommand \url  [0]{\begingroup\@sanitize@url \@url }%
\providecommand \@url [1]{\endgroup\@href {#1}{\urlprefix }}%
\providecommand \urlprefix  [0]{URL }%
\providecommand \Eprint [0]{\href }%
\providecommand \doibase [0]{https://doi.org/}%
\providecommand \selectlanguage [0]{\@gobble}%
\providecommand \bibinfo  [0]{\@secondoftwo}%
\providecommand \bibfield  [0]{\@secondoftwo}%
\providecommand \translation [1]{[#1]}%
\providecommand \BibitemOpen [0]{}%
\providecommand \bibitemStop [0]{}%
\providecommand \bibitemNoStop [0]{.\EOS\space}%
\providecommand \EOS [0]{\spacefactor3000\relax}%
\providecommand \BibitemShut  [1]{\csname bibitem#1\endcsname}%
\let\auto@bib@innerbib\@empty
\bibitem [{\citenamefont {Wallraff}\ \emph {et~al.}(2004)\citenamefont
  {Wallraff}, \citenamefont {Schuster}, \citenamefont {Blais}, \citenamefont
  {Frunzio}, \citenamefont {Huang}, \citenamefont {Majer}, \citenamefont
  {Kumar}, \citenamefont {Girvin},\ and\ \citenamefont
  {Schoelkopf}}]{Wallraff2004SCQ}%
  \BibitemOpen
  \bibfield  {author} {\bibinfo {author} {\bibfnamefont {A.}~\bibnamefont
  {Wallraff}}, \bibinfo {author} {\bibfnamefont {D.~I.}\ \bibnamefont
  {Schuster}}, \bibinfo {author} {\bibfnamefont {A.}~\bibnamefont {Blais}},
  \bibinfo {author} {\bibfnamefont {L.}~\bibnamefont {Frunzio}}, \bibinfo
  {author} {\bibfnamefont {R.~S.}\ \bibnamefont {Huang}}, \bibinfo {author}
  {\bibfnamefont {J.}~\bibnamefont {Majer}}, \bibinfo {author} {\bibfnamefont
  {S.}~\bibnamefont {Kumar}}, \bibinfo {author} {\bibfnamefont {S.~M.}\
  \bibnamefont {Girvin}},\ and\ \bibinfo {author} {\bibfnamefont {R.~J.}\
  \bibnamefont {Schoelkopf}},\ }\bibfield  {title} {\bibinfo {title} {Strong
  coupling of a single photon to a superconducting qubit using circuit quantum
  electrodynamics},\ }\href {https://doi.org/10} {\bibfield  {journal}
  {\bibinfo  {journal} {Nature}\ }\textbf {\bibinfo {volume} {431}},\ \bibinfo
  {pages} {162} (\bibinfo {year} {2004})}\BibitemShut {NoStop}%
\bibitem [{\citenamefont {You}\ and\ \citenamefont
  {Nori}(2011)}]{You2011SCQrev}%
  \BibitemOpen
  \bibfield  {author} {\bibinfo {author} {\bibfnamefont {J.~Q.}\ \bibnamefont
  {You}}\ and\ \bibinfo {author} {\bibfnamefont {F.}~\bibnamefont {Nori}},\
  }\bibfield  {title} {\bibinfo {title} {Atomic physics and quantum optics
  using superconducting circuits},\ }\href {https://doi.org/10} {\bibfield
  {journal} {\bibinfo  {journal} {Nature}\ }\textbf {\bibinfo {volume} {474}},\
  \bibinfo {pages} {589} (\bibinfo {year} {2011})}\BibitemShut {NoStop}%
\bibitem [{\citenamefont {Jelezko}\ \emph {et~al.}(2004)\citenamefont
  {Jelezko}, \citenamefont {Gaebel}, \citenamefont {Popa}, \citenamefont
  {Gruber},\ and\ \citenamefont {Wrachtrup}}]{Jelezko2004eRabi}%
  \BibitemOpen
  \bibfield  {author} {\bibinfo {author} {\bibfnamefont {F.}~\bibnamefont
  {Jelezko}}, \bibinfo {author} {\bibfnamefont {T.}~\bibnamefont {Gaebel}},
  \bibinfo {author} {\bibfnamefont {I.}~\bibnamefont {Popa}}, \bibinfo {author}
  {\bibfnamefont {A.}~\bibnamefont {Gruber}},\ and\ \bibinfo {author}
  {\bibfnamefont {J.}~\bibnamefont {Wrachtrup}},\ }\bibfield  {title} {\bibinfo
  {title} {Observation of coherent oscillations in a single electron spin},\
  }\href {https://doi.org/10.1103/PhysRevLett.92.076401} {\bibfield  {journal}
  {\bibinfo  {journal} {Phys. Rev. Lett.}\ }\textbf {\bibinfo {volume} {92}},\
  \bibinfo {pages} {076401} (\bibinfo {year} {2004})}\BibitemShut {NoStop}%
\bibitem [{\citenamefont {Koppens}\ \emph {et~al.}(2006)\citenamefont
  {Koppens}, \citenamefont {Buizert}, \citenamefont {Tielrooij}, \citenamefont
  {Vink}, \citenamefont {Nowack}, \citenamefont {Meunier}, \citenamefont
  {Kouwenhoven},\ and\ \citenamefont {Vandersypen}}]{Vandersypen2006Qdot}%
  \BibitemOpen
  \bibfield  {author} {\bibinfo {author} {\bibfnamefont {F.~H.~L.}\
  \bibnamefont {Koppens}}, \bibinfo {author} {\bibfnamefont {C.}~\bibnamefont
  {Buizert}}, \bibinfo {author} {\bibfnamefont {K.~J.}\ \bibnamefont
  {Tielrooij}}, \bibinfo {author} {\bibfnamefont {I.~T.}\ \bibnamefont {Vink}},
  \bibinfo {author} {\bibfnamefont {K.~C.}\ \bibnamefont {Nowack}}, \bibinfo
  {author} {\bibfnamefont {T.}~\bibnamefont {Meunier}}, \bibinfo {author}
  {\bibfnamefont {L.~P.}\ \bibnamefont {Kouwenhoven}},\ and\ \bibinfo {author}
  {\bibfnamefont {L.~M.~K.}\ \bibnamefont {Vandersypen}},\ }\bibfield  {title}
  {\bibinfo {title} {Driven coherent oscillations of a single electron spin in
  a quantum dot},\ }\href {https://doi.org/10} {\bibfield  {journal} {\bibinfo
  {journal} {Nature}\ }\textbf {\bibinfo {volume} {442}},\ \bibinfo {pages}
  {766} (\bibinfo {year} {2006})}\BibitemShut {NoStop}%
\bibitem [{\citenamefont {Casola}\ \emph {et~al.}(2018)\citenamefont {Casola},
  \citenamefont {van~der Sar},\ and\ \citenamefont
  {Yacoby}}]{Yacoby_RevNVCondMat2018}%
  \BibitemOpen
  \bibfield  {author} {\bibinfo {author} {\bibfnamefont {F.}~\bibnamefont
  {Casola}}, \bibinfo {author} {\bibfnamefont {T.}~\bibnamefont {van~der
  Sar}},\ and\ \bibinfo {author} {\bibfnamefont {A.}~\bibnamefont {Yacoby}},\
  }\bibfield  {title} {\bibinfo {title} {Probing condensed matter physics with
  magnetometry based on nitrogen-vacancy centres in diamond},\ }\href
  {https://doi.org/10.1038/natrevmats.2017.88} {\bibfield  {journal} {\bibinfo
  {journal} {Nature Reviews Materials}\ }\textbf {\bibinfo {volume} {3}},\
  \bibinfo {pages} {17088} (\bibinfo {year} {2018})}\BibitemShut {NoStop}%
\bibitem [{\citenamefont {Black}\ \emph {et~al.}(1995)\citenamefont {Black},
  \citenamefont {Wellstood}, \citenamefont {Dantsker}, \citenamefont {Miklich},
  \citenamefont {Koelle}, \citenamefont {Ludwig},\ and\ \citenamefont
  {Clarke}}]{Clarke_RFSQUID1995}%
  \BibitemOpen
  \bibfield  {author} {\bibinfo {author} {\bibfnamefont {R.~C.}\ \bibnamefont
  {Black}}, \bibinfo {author} {\bibfnamefont {F.~C.}\ \bibnamefont
  {Wellstood}}, \bibinfo {author} {\bibfnamefont {E.}~\bibnamefont {Dantsker}},
  \bibinfo {author} {\bibfnamefont {A.~H.}\ \bibnamefont {Miklich}}, \bibinfo
  {author} {\bibfnamefont {D.}~\bibnamefont {Koelle}}, \bibinfo {author}
  {\bibfnamefont {F.}~\bibnamefont {Ludwig}},\ and\ \bibinfo {author}
  {\bibfnamefont {J.}~\bibnamefont {Clarke}},\ }\bibfield  {title} {\bibinfo
  {title} {Imaging radio-frequency fields using a scanning squid microscope},\
  }\href {https://doi.org/10.1063/1.113258} {\bibfield  {journal} {\bibinfo
  {journal} {Applied Physics Letters}\ }\textbf {\bibinfo {volume} {66}},\
  \bibinfo {pages} {1267} (\bibinfo {year} {1995})}\BibitemShut {NoStop}%
\bibitem [{\citenamefont {Agrawal}\ \emph {et~al.}(1997)\citenamefont
  {Agrawal}, \citenamefont {Neuzil},\ and\ \citenamefont {van~der
  Weide}}]{vanderWeide_RFtip1997}%
  \BibitemOpen
  \bibfield  {author} {\bibinfo {author} {\bibfnamefont {V.}~\bibnamefont
  {Agrawal}}, \bibinfo {author} {\bibfnamefont {P.}~\bibnamefont {Neuzil}},\
  and\ \bibinfo {author} {\bibfnamefont {D.~W.}\ \bibnamefont {van~der
  Weide}},\ }\bibfield  {title} {\bibinfo {title} {{A microfabricated tip for
  simultaneous acquisition of sample topography and high-frequency magnetic
  field}},\ }\href {https://doi.org/10.1063/1.120073} {\bibfield  {journal}
  {\bibinfo  {journal} {Applied Physics Letters}\ }\textbf {\bibinfo {volume}
  {71}},\ \bibinfo {pages} {2343} (\bibinfo {year} {1997})}\BibitemShut
  {NoStop}%
\bibitem [{\citenamefont {Lee}\ \emph {et~al.}(2000)\citenamefont {Lee},
  \citenamefont {Vlahacos}, \citenamefont {Feenstra}, \citenamefont {Schwartz},
  \citenamefont {Steinhauer}, \citenamefont {Wellstood},\ and\ \citenamefont
  {Anlage}}]{Lee_MWmicroscope2000}%
  \BibitemOpen
  \bibfield  {author} {\bibinfo {author} {\bibfnamefont {S.-C.}\ \bibnamefont
  {Lee}}, \bibinfo {author} {\bibfnamefont {C.~P.}\ \bibnamefont {Vlahacos}},
  \bibinfo {author} {\bibfnamefont {B.~J.}\ \bibnamefont {Feenstra}}, \bibinfo
  {author} {\bibfnamefont {A.}~\bibnamefont {Schwartz}}, \bibinfo {author}
  {\bibfnamefont {D.~E.}\ \bibnamefont {Steinhauer}}, \bibinfo {author}
  {\bibfnamefont {F.~C.}\ \bibnamefont {Wellstood}},\ and\ \bibinfo {author}
  {\bibfnamefont {S.~M.}\ \bibnamefont {Anlage}},\ }\bibfield  {title}
  {\bibinfo {title} {{Magnetic permeability imaging of metals with a scanning
  near-field microwave microscope}},\ }\href
  {https://doi.org/10.1063/1.1332978} {\bibfield  {journal} {\bibinfo
  {journal} {Applied Physics Letters}\ }\textbf {\bibinfo {volume} {77}},\
  \bibinfo {pages} {4404} (\bibinfo {year} {2000})}\BibitemShut {NoStop}%
\bibitem [{\citenamefont {B{\"o}hi}\ \emph {et~al.}(2010)\citenamefont
  {B{\"o}hi}, \citenamefont {Riedel}, \citenamefont {H{\"a}nsch},\ and\
  \citenamefont {Treutlein}}]{Boehi_MWimgAtoms2010}%
  \BibitemOpen
  \bibfield  {author} {\bibinfo {author} {\bibfnamefont {P.}~\bibnamefont
  {B{\"o}hi}}, \bibinfo {author} {\bibfnamefont {M.~F.}\ \bibnamefont
  {Riedel}}, \bibinfo {author} {\bibfnamefont {T.~W.}\ \bibnamefont
  {H{\"a}nsch}},\ and\ \bibinfo {author} {\bibfnamefont {P.}~\bibnamefont
  {Treutlein}},\ }\bibfield  {title} {\bibinfo {title} {{Imaging of microwave
  fields using ultracold atoms}},\ }\href {https://doi.org/10.1063/1.3470591}
  {\bibfield  {journal} {\bibinfo  {journal} {Applied Physics Letters}\
  }\textbf {\bibinfo {volume} {97}},\ \bibinfo {pages} {051101} (\bibinfo
  {year} {2010})}\BibitemShut {NoStop}%
\bibitem [{\citenamefont {B{\"o}hi}\ and\ \citenamefont
  {Treutlein}(2012)}]{Boehi_MWimgVapor2012}%
  \BibitemOpen
  \bibfield  {author} {\bibinfo {author} {\bibfnamefont {P.}~\bibnamefont
  {B{\"o}hi}}\ and\ \bibinfo {author} {\bibfnamefont {P.}~\bibnamefont
  {Treutlein}},\ }\bibfield  {title} {\bibinfo {title} {{Simple microwave field
  imaging technique using hot atomic vapor cells}},\ }\href
  {https://doi.org/10.1063/1.4760267} {\bibfield  {journal} {\bibinfo
  {journal} {Applied Physics Letters}\ }\textbf {\bibinfo {volume} {101}},\
  \bibinfo {pages} {181107} (\bibinfo {year} {2012})}\BibitemShut {NoStop}%
\bibitem [{\citenamefont {Ockeloen}\ \emph {et~al.}(2013)\citenamefont
  {Ockeloen}, \citenamefont {Schmied}, \citenamefont {Riedel},\ and\
  \citenamefont {Treutlein}}]{Ockeloen_MW_BECmag2013}%
  \BibitemOpen
  \bibfield  {author} {\bibinfo {author} {\bibfnamefont {C.~F.}\ \bibnamefont
  {Ockeloen}}, \bibinfo {author} {\bibfnamefont {R.}~\bibnamefont {Schmied}},
  \bibinfo {author} {\bibfnamefont {M.~F.}\ \bibnamefont {Riedel}},\ and\
  \bibinfo {author} {\bibfnamefont {P.}~\bibnamefont {Treutlein}},\ }\bibfield
  {title} {\bibinfo {title} {Quantum metrology with a scanning probe atom
  interferometer},\ }\href {https://doi.org/10.1103/PhysRevLett.111.143001}
  {\bibfield  {journal} {\bibinfo  {journal} {Phys. Rev. Lett.}\ }\textbf
  {\bibinfo {volume} {111}},\ \bibinfo {pages} {143001} (\bibinfo {year}
  {2013})}\BibitemShut {NoStop}%
\bibitem [{\citenamefont {van~der Sar}\ \emph {et~al.}(2015)\citenamefont
  {van~der Sar}, \citenamefont {Casola}, \citenamefont {Walsworth},\ and\
  \citenamefont {Yacoby}}]{vanderSar_SpinWaves2015}%
  \BibitemOpen
  \bibfield  {author} {\bibinfo {author} {\bibfnamefont {T.}~\bibnamefont
  {van~der Sar}}, \bibinfo {author} {\bibfnamefont {F.}~\bibnamefont {Casola}},
  \bibinfo {author} {\bibfnamefont {R.}~\bibnamefont {Walsworth}},\ and\
  \bibinfo {author} {\bibfnamefont {A.}~\bibnamefont {Yacoby}},\ }\bibfield
  {title} {\bibinfo {title} {Nanometre-scale probing of spin waves using single
  electron spins},\ }\href {https://doi.org/10.1038/ncomms8886} {\bibfield
  {journal} {\bibinfo  {journal} {Nature Communications}\ }\textbf {\bibinfo
  {volume} {6}},\ \bibinfo {pages} {7886} (\bibinfo {year} {2015})}\BibitemShut
  {NoStop}%
\bibitem [{\citenamefont {Appel}\ \emph {et~al.}(2015)\citenamefont {Appel},
  \citenamefont {Ganzhorn}, \citenamefont {Neu},\ and\ \citenamefont
  {Maletinsky}}]{Maletinsky_MWsens2015njp}%
  \BibitemOpen
  \bibfield  {author} {\bibinfo {author} {\bibfnamefont {P.}~\bibnamefont
  {Appel}}, \bibinfo {author} {\bibfnamefont {M.}~\bibnamefont {Ganzhorn}},
  \bibinfo {author} {\bibfnamefont {E.}~\bibnamefont {Neu}},\ and\ \bibinfo
  {author} {\bibfnamefont {P.}~\bibnamefont {Maletinsky}},\ }\bibfield  {title}
  {\bibinfo {title} {Nanoscale microwave imaging with a single electron spin in
  diamond},\ }\href {https://doi.org/10.1088/1367-2630/17/11/112001} {\bibfield
   {journal} {\bibinfo  {journal} {New Journal of Physics}\ }\textbf {\bibinfo
  {volume} {17}},\ \bibinfo {pages} {112001} (\bibinfo {year}
  {2015})}\BibitemShut {NoStop}%
\bibitem [{\citenamefont {Rao}\ and\ \citenamefont
  {Suter}(2020)}]{Koti2020_LAC}%
  \BibitemOpen
  \bibfield  {author} {\bibinfo {author} {\bibfnamefont {K.~R.~K.}\
  \bibnamefont {Rao}}\ and\ \bibinfo {author} {\bibfnamefont {D.}~\bibnamefont
  {Suter}},\ }\bibfield  {title} {\bibinfo {title} {Level anti-crossings of a
  nitrogen-vacancy center in diamond: decoherence-free subspaces and 3d sensors
  of microwave magnetic fields},\ }\href
  {https://doi.org/10.1088/1367-2630/abc083} {\bibfield  {journal} {\bibinfo
  {journal} {New Journal of Physics}\ }\textbf {\bibinfo {volume} {22}},\
  \bibinfo {pages} {103065} (\bibinfo {year} {2020})}\BibitemShut {NoStop}%
\bibitem [{\citenamefont {Doherty}\ \emph {et~al.}(2013)\citenamefont
  {Doherty}, \citenamefont {Manson}, \citenamefont {Delaney}, \citenamefont
  {Jelezko}, \citenamefont {Wrachtrup},\ and\ \citenamefont
  {Hollenberg}}]{RevDoherty}%
  \BibitemOpen
  \bibfield  {author} {\bibinfo {author} {\bibfnamefont {M.~W.}\ \bibnamefont
  {Doherty}}, \bibinfo {author} {\bibfnamefont {N.~B.}\ \bibnamefont {Manson}},
  \bibinfo {author} {\bibfnamefont {P.}~\bibnamefont {Delaney}}, \bibinfo
  {author} {\bibfnamefont {F.}~\bibnamefont {Jelezko}}, \bibinfo {author}
  {\bibfnamefont {J.}~\bibnamefont {Wrachtrup}},\ and\ \bibinfo {author}
  {\bibfnamefont {L.~C.}\ \bibnamefont {Hollenberg}},\ }\bibfield  {title}
  {\bibinfo {title} {The nitrogen-vacancy colour centre in diamond},\ }\href
  {https://doi.org/http://dx.doi.org/10.1016/j.physrep.2013.02.001} {\bibfield
  {journal} {\bibinfo  {journal} {Physics Reports}\ }\textbf {\bibinfo {volume}
  {528}},\ \bibinfo {pages} {1 } (\bibinfo {year} {2013})}\BibitemShut
  {NoStop}%
\bibitem [{\citenamefont {Suter}\ and\ \citenamefont
  {Jelezko}(2017)}]{SuterNVRev2017}%
  \BibitemOpen
  \bibfield  {author} {\bibinfo {author} {\bibfnamefont {D.}~\bibnamefont
  {Suter}}\ and\ \bibinfo {author} {\bibfnamefont {F.}~\bibnamefont
  {Jelezko}},\ }\bibfield  {title} {\bibinfo {title} {Single-spin magnetic
  resonance in the nitrogen-vacancy center of diamond},\ }\href
  {https://doi.org/https://doi.org/10.1016/j.pnmrs.2016.12.001} {\bibfield
  {journal} {\bibinfo  {journal} {Progress in Nuclear Magnetic Resonance
  Spectroscopy}\ }\textbf {\bibinfo {volume} {98-99}},\ \bibinfo {pages} {50 }
  (\bibinfo {year} {2017})}\BibitemShut {NoStop}%
\bibitem [{\citenamefont {Schirhagl}\ \emph {et~al.}(2014)\citenamefont
  {Schirhagl}, \citenamefont {Chang}, \citenamefont {Loretz},\ and\
  \citenamefont {Degen}}]{RevDegen}%
  \BibitemOpen
  \bibfield  {author} {\bibinfo {author} {\bibfnamefont {R.}~\bibnamefont
  {Schirhagl}}, \bibinfo {author} {\bibfnamefont {K.}~\bibnamefont {Chang}},
  \bibinfo {author} {\bibfnamefont {M.}~\bibnamefont {Loretz}},\ and\ \bibinfo
  {author} {\bibfnamefont {C.~L.}\ \bibnamefont {Degen}},\ }\bibfield  {title}
  {\bibinfo {title} {Nitrogen-vacancy centers in diamond: Nanoscale sensors for
  physics and biology},\ }\href
  {https://doi.org/10.1146/annurev-physchem-040513-103659} {\bibfield
  {journal} {\bibinfo  {journal} {Annual Review of Physical Chemistry}\
  }\textbf {\bibinfo {volume} {65}},\ \bibinfo {pages} {83} (\bibinfo {year}
  {2014})}\BibitemShut {NoStop}%
\bibitem [{\citenamefont {Rondin}\ \emph {et~al.}(2014)\citenamefont {Rondin},
  \citenamefont {Tetienne}, \citenamefont {Hingant}, \citenamefont {Roch},
  \citenamefont {Maletinsky},\ and\ \citenamefont {Jacques}}]{RevJacques}%
  \BibitemOpen
  \bibfield  {author} {\bibinfo {author} {\bibfnamefont {L.}~\bibnamefont
  {Rondin}}, \bibinfo {author} {\bibfnamefont {J.-P.}\ \bibnamefont
  {Tetienne}}, \bibinfo {author} {\bibfnamefont {T.}~\bibnamefont {Hingant}},
  \bibinfo {author} {\bibfnamefont {J.-F.}\ \bibnamefont {Roch}}, \bibinfo
  {author} {\bibfnamefont {P.}~\bibnamefont {Maletinsky}},\ and\ \bibinfo
  {author} {\bibfnamefont {V.}~\bibnamefont {Jacques}},\ }\bibfield  {title}
  {\bibinfo {title} {Magnetometry with nitrogen-vacancy defects in diamond},\
  }\href {http://stacks.iop.org/0034-4885/77/i=5/a=056503} {\bibfield
  {journal} {\bibinfo  {journal} {Reports on Progress in Physics}\ }\textbf
  {\bibinfo {volume} {77}},\ \bibinfo {pages} {056503} (\bibinfo {year}
  {2014})}\BibitemShut {NoStop}%
\bibitem [{\citenamefont {Hong}\ \emph {et~al.}(2013)\citenamefont {Hong},
  \citenamefont {Grinolds}, \citenamefont {Pham}, \citenamefont {Sage},
  \citenamefont {Luan}, \citenamefont {Walsworth},\ and\ \citenamefont
  {Yacoby}}]{RevWalsworth}%
  \BibitemOpen
  \bibfield  {author} {\bibinfo {author} {\bibfnamefont {S.}~\bibnamefont
  {Hong}}, \bibinfo {author} {\bibfnamefont {M.~S.}\ \bibnamefont {Grinolds}},
  \bibinfo {author} {\bibfnamefont {L.~M.}\ \bibnamefont {Pham}}, \bibinfo
  {author} {\bibfnamefont {D.~L.}\ \bibnamefont {Sage}}, \bibinfo {author}
  {\bibfnamefont {L.}~\bibnamefont {Luan}}, \bibinfo {author} {\bibfnamefont
  {R.~L.}\ \bibnamefont {Walsworth}},\ and\ \bibinfo {author} {\bibfnamefont
  {A.}~\bibnamefont {Yacoby}},\ }\bibfield  {title} {\bibinfo {title}
  {Nanoscale magnetometry with nv centers in diamond},\ }\href
  {https://doi.org/10.1557/mrs.2013.23} {\bibfield  {journal} {\bibinfo
  {journal} {MRS Bulletin}\ }\textbf {\bibinfo {volume} {38}},\ \bibinfo
  {pages} {155} (\bibinfo {year} {2013})}\BibitemShut {NoStop}%
\bibitem [{\citenamefont {Childress}\ and\ \citenamefont
  {Hanson}(2013)}]{RevChildress}%
  \BibitemOpen
  \bibfield  {author} {\bibinfo {author} {\bibfnamefont {L.}~\bibnamefont
  {Childress}}\ and\ \bibinfo {author} {\bibfnamefont {R.}~\bibnamefont
  {Hanson}},\ }\bibfield  {title} {\bibinfo {title} {Diamond nv centers for
  quantum computing and quantum networks},\ }\href
  {https://doi.org/10.1557/mrs.2013.20} {\bibfield  {journal} {\bibinfo
  {journal} {MRS Bulletin}\ }\textbf {\bibinfo {volume} {38}},\ \bibinfo
  {pages} {134} (\bibinfo {year} {2013})}\BibitemShut {NoStop}%
\bibitem [{\citenamefont {Dolde}\ \emph {et~al.}(2011)\citenamefont {Dolde},
  \citenamefont {Fedder}, \citenamefont {Doherty}, \citenamefont {Nobauer},
  \citenamefont {Rempp}, \citenamefont {Balasubramanian}, \citenamefont {Wolf},
  \citenamefont {Reinhard}, \citenamefont {Hollenberg}, \citenamefont
  {Jelezko},\ and\ \citenamefont {Wrachtrup}}]{WrachtrupESens}%
  \BibitemOpen
  \bibfield  {author} {\bibinfo {author} {\bibfnamefont {F.}~\bibnamefont
  {Dolde}}, \bibinfo {author} {\bibfnamefont {H.}~\bibnamefont {Fedder}},
  \bibinfo {author} {\bibfnamefont {M.~W.}\ \bibnamefont {Doherty}}, \bibinfo
  {author} {\bibfnamefont {T.}~\bibnamefont {Nobauer}}, \bibinfo {author}
  {\bibfnamefont {F.}~\bibnamefont {Rempp}}, \bibinfo {author} {\bibfnamefont
  {G.}~\bibnamefont {Balasubramanian}}, \bibinfo {author} {\bibfnamefont
  {T.}~\bibnamefont {Wolf}}, \bibinfo {author} {\bibfnamefont {F.}~\bibnamefont
  {Reinhard}}, \bibinfo {author} {\bibfnamefont {L.~C.~L.}\ \bibnamefont
  {Hollenberg}}, \bibinfo {author} {\bibfnamefont {F.}~\bibnamefont
  {Jelezko}},\ and\ \bibinfo {author} {\bibfnamefont {J.}~\bibnamefont
  {Wrachtrup}},\ }\bibfield  {title} {\bibinfo {title} {Electric-field sensing
  using single diamond spins},\ }\href {https://doi.org/10.1038/nphys1969}
  {\bibfield  {journal} {\bibinfo  {journal} {Nature Physics}\ }\textbf
  {\bibinfo {volume} {7}},\ \bibinfo {pages} {459} (\bibinfo {year}
  {2011})}\BibitemShut {NoStop}%
\bibitem [{\citenamefont {Balasubramanian}\ \emph {et~al.}(2008)\citenamefont
  {Balasubramanian}, \citenamefont {Chan}, \citenamefont {Kolesov},
  \citenamefont {Al-Hmoud}, \citenamefont {Tisler}, \citenamefont {Shin},
  \citenamefont {Kim}, \citenamefont {Wojcik}, \citenamefont {Hemmer},
  \citenamefont {Krueger} \emph {et~al.}}]{JW2008Nat}%
  \BibitemOpen
  \bibfield  {author} {\bibinfo {author} {\bibfnamefont {G.}~\bibnamefont
  {Balasubramanian}}, \bibinfo {author} {\bibfnamefont {I.}~\bibnamefont
  {Chan}}, \bibinfo {author} {\bibfnamefont {R.}~\bibnamefont {Kolesov}},
  \bibinfo {author} {\bibfnamefont {M.}~\bibnamefont {Al-Hmoud}}, \bibinfo
  {author} {\bibfnamefont {J.}~\bibnamefont {Tisler}}, \bibinfo {author}
  {\bibfnamefont {C.}~\bibnamefont {Shin}}, \bibinfo {author} {\bibfnamefont
  {C.}~\bibnamefont {Kim}}, \bibinfo {author} {\bibfnamefont {A.}~\bibnamefont
  {Wojcik}}, \bibinfo {author} {\bibfnamefont {P.~R.}\ \bibnamefont {Hemmer}},
  \bibinfo {author} {\bibfnamefont {A.}~\bibnamefont {Krueger}}, \emph
  {et~al.},\ }\bibfield  {title} {\bibinfo {title} {Nanoscale imaging
  magnetometry with diamond spins under ambient conditions},\ }\href
  {https://doi.org/10.1038/nature07278} {\bibfield  {journal} {\bibinfo
  {journal} {Nature}\ }\textbf {\bibinfo {volume} {455}},\ \bibinfo {pages}
  {648} (\bibinfo {year} {2008})}\BibitemShut {NoStop}%
\bibitem [{\citenamefont {Degen}(2008)}]{Degen_Mag2008APL}%
  \BibitemOpen
  \bibfield  {author} {\bibinfo {author} {\bibfnamefont {C.~L.}\ \bibnamefont
  {Degen}},\ }\bibfield  {title} {\bibinfo {title} {{Scanning magnetic field
  microscope with a diamond single-spin sensor}},\ }\href
  {https://doi.org/10.1063/1.2943282} {\bibfield  {journal} {\bibinfo
  {journal} {Applied Physics Letters}\ }\textbf {\bibinfo {volume} {92}},\
  \bibinfo {pages} {243111} (\bibinfo {year} {2008})}\BibitemShut {NoStop}%
\bibitem [{\citenamefont {Maze}\ \emph {et~al.}(2008)\citenamefont {Maze},
  \citenamefont {Stanwix}, \citenamefont {Hodges}, \citenamefont {Hong},
  \citenamefont {Taylor}, \citenamefont {Cappellaro}, \citenamefont {Jiang},
  \citenamefont {Dutt}, \citenamefont {Togan}, \citenamefont {Zibrov},
  \citenamefont {Yacoby}, \citenamefont {Walsworth},\ and\ \citenamefont
  {Lukin}}]{Lukin_Mag2008Nat}%
  \BibitemOpen
  \bibfield  {author} {\bibinfo {author} {\bibfnamefont {J.~R.}\ \bibnamefont
  {Maze}}, \bibinfo {author} {\bibfnamefont {P.~L.}\ \bibnamefont {Stanwix}},
  \bibinfo {author} {\bibfnamefont {J.~S.}\ \bibnamefont {Hodges}}, \bibinfo
  {author} {\bibfnamefont {S.}~\bibnamefont {Hong}}, \bibinfo {author}
  {\bibfnamefont {J.~M.}\ \bibnamefont {Taylor}}, \bibinfo {author}
  {\bibfnamefont {P.}~\bibnamefont {Cappellaro}}, \bibinfo {author}
  {\bibfnamefont {L.}~\bibnamefont {Jiang}}, \bibinfo {author} {\bibfnamefont
  {M.~V.~G.}\ \bibnamefont {Dutt}}, \bibinfo {author} {\bibfnamefont
  {E.}~\bibnamefont {Togan}}, \bibinfo {author} {\bibfnamefont {A.~S.}\
  \bibnamefont {Zibrov}}, \bibinfo {author} {\bibfnamefont {A.}~\bibnamefont
  {Yacoby}}, \bibinfo {author} {\bibfnamefont {R.~L.}\ \bibnamefont
  {Walsworth}},\ and\ \bibinfo {author} {\bibfnamefont {M.~D.}\ \bibnamefont
  {Lukin}},\ }\bibfield  {title} {\bibinfo {title} {Nanoscale magnetic sensing
  with an individual electronic spin in diamond},\ }\href
  {https://doi.org/10.1038/nature07279} {\bibfield  {journal} {\bibinfo
  {journal} {Nature}\ }\textbf {\bibinfo {volume} {455}},\ \bibinfo {pages}
  {644} (\bibinfo {year} {2008})}\BibitemShut {NoStop}%
\bibitem [{\citenamefont {Staudacher}\ \emph {et~al.}(2013)\citenamefont
  {Staudacher}, \citenamefont {Shi}, \citenamefont {Pezzagna}, \citenamefont
  {Meijer}, \citenamefont {Du}, \citenamefont {Meriles}, \citenamefont
  {Reinhard},\ and\ \citenamefont {Wrachtrup}}]{Staudacher2013}%
  \BibitemOpen
  \bibfield  {author} {\bibinfo {author} {\bibfnamefont {T.}~\bibnamefont
  {Staudacher}}, \bibinfo {author} {\bibfnamefont {F.}~\bibnamefont {Shi}},
  \bibinfo {author} {\bibfnamefont {S.}~\bibnamefont {Pezzagna}}, \bibinfo
  {author} {\bibfnamefont {J.}~\bibnamefont {Meijer}}, \bibinfo {author}
  {\bibfnamefont {J.}~\bibnamefont {Du}}, \bibinfo {author} {\bibfnamefont
  {C.~A.}\ \bibnamefont {Meriles}}, \bibinfo {author} {\bibfnamefont
  {F.}~\bibnamefont {Reinhard}},\ and\ \bibinfo {author} {\bibfnamefont
  {J.}~\bibnamefont {Wrachtrup}},\ }\bibfield  {title} {\bibinfo {title}
  {Nuclear magnetic resonance spectroscopy on a (5-nanometer)3 sample volume},\
  }\href {https://doi.org/10.1126/science.1231675} {\bibfield  {journal}
  {\bibinfo  {journal} {Science}\ }\textbf {\bibinfo {volume} {339}},\ \bibinfo
  {pages} {561} (\bibinfo {year} {2013})}\BibitemShut {NoStop}%
\bibitem [{\citenamefont {Boss}\ \emph {et~al.}(2017)\citenamefont {Boss},
  \citenamefont {Cujia}, \citenamefont {Zopes},\ and\ \citenamefont
  {Degen}}]{DegenScience2017}%
  \BibitemOpen
  \bibfield  {author} {\bibinfo {author} {\bibfnamefont {J.~M.}\ \bibnamefont
  {Boss}}, \bibinfo {author} {\bibfnamefont {K.~S.}\ \bibnamefont {Cujia}},
  \bibinfo {author} {\bibfnamefont {J.}~\bibnamefont {Zopes}},\ and\ \bibinfo
  {author} {\bibfnamefont {C.~L.}\ \bibnamefont {Degen}},\ }\bibfield  {title}
  {\bibinfo {title} {Quantum sensing with arbitrary frequency resolution},\
  }\href {https://doi.org/10.1126/science.aam7009} {\bibfield  {journal}
  {\bibinfo  {journal} {Science}\ }\textbf {\bibinfo {volume} {356}},\ \bibinfo
  {pages} {837} (\bibinfo {year} {2017})}\BibitemShut {NoStop}%
\bibitem [{\citenamefont {Schmitt}\ \emph {et~al.}(2017)\citenamefont
  {Schmitt}, \citenamefont {Gefen}, \citenamefont {St{\"u}rner}, \citenamefont
  {Unden}, \citenamefont {Wolff}, \citenamefont {M{\"u}ller}, \citenamefont
  {Scheuer}, \citenamefont {Naydenov}, \citenamefont {Markham}, \citenamefont
  {Pezzagna}, \citenamefont {Meijer}, \citenamefont {Schwarz}, \citenamefont
  {Plenio}, \citenamefont {Retzker}, \citenamefont {McGuinness},\ and\
  \citenamefont {Jelezko}}]{JelezkoScience2017}%
  \BibitemOpen
  \bibfield  {author} {\bibinfo {author} {\bibfnamefont {S.}~\bibnamefont
  {Schmitt}}, \bibinfo {author} {\bibfnamefont {T.}~\bibnamefont {Gefen}},
  \bibinfo {author} {\bibfnamefont {F.~M.}\ \bibnamefont {St{\"u}rner}},
  \bibinfo {author} {\bibfnamefont {T.}~\bibnamefont {Unden}}, \bibinfo
  {author} {\bibfnamefont {G.}~\bibnamefont {Wolff}}, \bibinfo {author}
  {\bibfnamefont {C.}~\bibnamefont {M{\"u}ller}}, \bibinfo {author}
  {\bibfnamefont {J.}~\bibnamefont {Scheuer}}, \bibinfo {author} {\bibfnamefont
  {B.}~\bibnamefont {Naydenov}}, \bibinfo {author} {\bibfnamefont
  {M.}~\bibnamefont {Markham}}, \bibinfo {author} {\bibfnamefont
  {S.}~\bibnamefont {Pezzagna}}, \bibinfo {author} {\bibfnamefont
  {J.}~\bibnamefont {Meijer}}, \bibinfo {author} {\bibfnamefont
  {I.}~\bibnamefont {Schwarz}}, \bibinfo {author} {\bibfnamefont
  {M.}~\bibnamefont {Plenio}}, \bibinfo {author} {\bibfnamefont
  {A.}~\bibnamefont {Retzker}}, \bibinfo {author} {\bibfnamefont {L.~P.}\
  \bibnamefont {McGuinness}},\ and\ \bibinfo {author} {\bibfnamefont
  {F.}~\bibnamefont {Jelezko}},\ }\bibfield  {title} {\bibinfo {title}
  {Submillihertz magnetic spectroscopy performed with a nanoscale quantum
  sensor},\ }\href {https://doi.org/10.1126/science.aam5532} {\bibfield
  {journal} {\bibinfo  {journal} {Science}\ }\textbf {\bibinfo {volume}
  {356}},\ \bibinfo {pages} {832} (\bibinfo {year} {2017})}\BibitemShut
  {NoStop}%
\bibitem [{\citenamefont {Maertz}\ \emph {et~al.}(2010)\citenamefont {Maertz},
  \citenamefont {Wijnheijmer}, \citenamefont {Fuchs}, \citenamefont
  {Nowakowski},\ and\ \citenamefont {Awschalom}}]{Awschalom_VecMag2010}%
  \BibitemOpen
  \bibfield  {author} {\bibinfo {author} {\bibfnamefont {B.~J.}\ \bibnamefont
  {Maertz}}, \bibinfo {author} {\bibfnamefont {A.~P.}\ \bibnamefont
  {Wijnheijmer}}, \bibinfo {author} {\bibfnamefont {G.~D.}\ \bibnamefont
  {Fuchs}}, \bibinfo {author} {\bibfnamefont {M.~E.}\ \bibnamefont
  {Nowakowski}},\ and\ \bibinfo {author} {\bibfnamefont {D.~D.}\ \bibnamefont
  {Awschalom}},\ }\bibfield  {title} {\bibinfo {title} {{Vector magnetic field
  microscopy using nitrogen vacancy centers in diamond}},\ }\href
  {https://doi.org/10.1063/1.3337096} {\bibfield  {journal} {\bibinfo
  {journal} {Applied Physics Letters}\ }\textbf {\bibinfo {volume} {96}},\
  \bibinfo {pages} {092504} (\bibinfo {year} {2010})}\BibitemShut {NoStop}%
\bibitem [{\citenamefont {Zhang}\ \emph {et~al.}(2018)\citenamefont {Zhang},
  \citenamefont {Yuan}, \citenamefont {Zhang}, \citenamefont {Xu},
  \citenamefont {Zhang}, \citenamefont {Li},\ and\ \citenamefont
  {Fang}}]{Fang_VecMag2018}%
  \BibitemOpen
  \bibfield  {author} {\bibinfo {author} {\bibfnamefont {C.}~\bibnamefont
  {Zhang}}, \bibinfo {author} {\bibfnamefont {H.}~\bibnamefont {Yuan}},
  \bibinfo {author} {\bibfnamefont {N.}~\bibnamefont {Zhang}}, \bibinfo
  {author} {\bibfnamefont {L.}~\bibnamefont {Xu}}, \bibinfo {author}
  {\bibfnamefont {J.}~\bibnamefont {Zhang}}, \bibinfo {author} {\bibfnamefont
  {B.}~\bibnamefont {Li}},\ and\ \bibinfo {author} {\bibfnamefont
  {J.}~\bibnamefont {Fang}},\ }\bibfield  {title} {\bibinfo {title} {Vector
  magnetometer based on synchronous manipulation of nitrogen-vacancy centers in
  all crystal directions},\ }\href {https://doi.org/10.1088/1361-6463/aab2d0}
  {\bibfield  {journal} {\bibinfo  {journal} {Journal of Physics D: Applied
  Physics}\ }\textbf {\bibinfo {volume} {51}},\ \bibinfo {pages} {155102}
  (\bibinfo {year} {2018})}\BibitemShut {NoStop}%
\bibitem [{\citenamefont {Clevenson}\ \emph {et~al.}(2018)\citenamefont
  {Clevenson}, \citenamefont {Pham}, \citenamefont {Teale}, \citenamefont
  {Johnson}, \citenamefont {Englund},\ and\ \citenamefont
  {Braje}}]{Clevenson_VecMag2018}%
  \BibitemOpen
  \bibfield  {author} {\bibinfo {author} {\bibfnamefont {H.}~\bibnamefont
  {Clevenson}}, \bibinfo {author} {\bibfnamefont {L.~M.}\ \bibnamefont {Pham}},
  \bibinfo {author} {\bibfnamefont {C.}~\bibnamefont {Teale}}, \bibinfo
  {author} {\bibfnamefont {K.}~\bibnamefont {Johnson}}, \bibinfo {author}
  {\bibfnamefont {D.}~\bibnamefont {Englund}},\ and\ \bibinfo {author}
  {\bibfnamefont {D.}~\bibnamefont {Braje}},\ }\bibfield  {title} {\bibinfo
  {title} {{Robust high-dynamic-range vector magnetometry with nitrogen-vacancy
  centers in diamond}},\ }\href {https://doi.org/10.1063/1.5034216} {\bibfield
  {journal} {\bibinfo  {journal} {Applied Physics Letters}\ }\textbf {\bibinfo
  {volume} {112}},\ \bibinfo {pages} {252406} (\bibinfo {year}
  {2018})}\BibitemShut {NoStop}%
\bibitem [{\citenamefont {Schloss}\ \emph {et~al.}(2018)\citenamefont
  {Schloss}, \citenamefont {Barry}, \citenamefont {Turner},\ and\ \citenamefont
  {Walsworth}}]{Walsworth_VecMag2018}%
  \BibitemOpen
  \bibfield  {author} {\bibinfo {author} {\bibfnamefont {J.~M.}\ \bibnamefont
  {Schloss}}, \bibinfo {author} {\bibfnamefont {J.~F.}\ \bibnamefont {Barry}},
  \bibinfo {author} {\bibfnamefont {M.~J.}\ \bibnamefont {Turner}},\ and\
  \bibinfo {author} {\bibfnamefont {R.~L.}\ \bibnamefont {Walsworth}},\
  }\bibfield  {title} {\bibinfo {title} {Simultaneous broadband vector
  magnetometry using solid-state spins},\ }\href
  {https://doi.org/10.1103/PhysRevApplied.10.034044} {\bibfield  {journal}
  {\bibinfo  {journal} {Phys. Rev. Appl.}\ }\textbf {\bibinfo {volume} {10}},\
  \bibinfo {pages} {034044} (\bibinfo {year} {2018})}\BibitemShut {NoStop}%
\bibitem [{\citenamefont {Zheng}\ \emph {et~al.}(2020)\citenamefont {Zheng},
  \citenamefont {Sun}, \citenamefont {Chatzidrosos}, \citenamefont {Zhang},
  \citenamefont {Nakamura}, \citenamefont {Sumiya}, \citenamefont {Ohshima},
  \citenamefont {Isoya}, \citenamefont {Wrachtrup}, \citenamefont
  {Wickenbrock},\ and\ \citenamefont {Budker}}]{Budker_VecMag2020}%
  \BibitemOpen
  \bibfield  {author} {\bibinfo {author} {\bibfnamefont {H.}~\bibnamefont
  {Zheng}}, \bibinfo {author} {\bibfnamefont {Z.}~\bibnamefont {Sun}}, \bibinfo
  {author} {\bibfnamefont {G.}~\bibnamefont {Chatzidrosos}}, \bibinfo {author}
  {\bibfnamefont {C.}~\bibnamefont {Zhang}}, \bibinfo {author} {\bibfnamefont
  {K.}~\bibnamefont {Nakamura}}, \bibinfo {author} {\bibfnamefont
  {H.}~\bibnamefont {Sumiya}}, \bibinfo {author} {\bibfnamefont
  {T.}~\bibnamefont {Ohshima}}, \bibinfo {author} {\bibfnamefont
  {J.}~\bibnamefont {Isoya}}, \bibinfo {author} {\bibfnamefont
  {J.}~\bibnamefont {Wrachtrup}}, \bibinfo {author} {\bibfnamefont
  {A.}~\bibnamefont {Wickenbrock}},\ and\ \bibinfo {author} {\bibfnamefont
  {D.}~\bibnamefont {Budker}},\ }\bibfield  {title} {\bibinfo {title}
  {Microwave-free vector magnetometry with nitrogen-vacancy centers along a
  single axis in diamond},\ }\href
  {https://doi.org/10.1103/PhysRevApplied.13.044023} {\bibfield  {journal}
  {\bibinfo  {journal} {Phys. Rev. Appl.}\ }\textbf {\bibinfo {volume} {13}},\
  \bibinfo {pages} {044023} (\bibinfo {year} {2020})}\BibitemShut {NoStop}%
\bibitem [{\citenamefont {Broadway}\ \emph {et~al.}(2020)\citenamefont
  {Broadway}, \citenamefont {Lillie}, \citenamefont {Scholten}, \citenamefont
  {Rohner}, \citenamefont {Dontschuk}, \citenamefont {Maletinsky},
  \citenamefont {Tetienne},\ and\ \citenamefont
  {Hollenberg}}]{Hollenberg_VecMag2020}%
  \BibitemOpen
  \bibfield  {author} {\bibinfo {author} {\bibfnamefont {D.}~\bibnamefont
  {Broadway}}, \bibinfo {author} {\bibfnamefont {S.}~\bibnamefont {Lillie}},
  \bibinfo {author} {\bibfnamefont {S.}~\bibnamefont {Scholten}}, \bibinfo
  {author} {\bibfnamefont {D.}~\bibnamefont {Rohner}}, \bibinfo {author}
  {\bibfnamefont {N.}~\bibnamefont {Dontschuk}}, \bibinfo {author}
  {\bibfnamefont {P.}~\bibnamefont {Maletinsky}}, \bibinfo {author}
  {\bibfnamefont {J.-P.}\ \bibnamefont {Tetienne}},\ and\ \bibinfo {author}
  {\bibfnamefont {L.}~\bibnamefont {Hollenberg}},\ }\bibfield  {title}
  {\bibinfo {title} {Improved current density and magnetization reconstruction
  through vector magnetic field measurements},\ }\href
  {https://doi.org/10.1103/PhysRevApplied.14.024076} {\bibfield  {journal}
  {\bibinfo  {journal} {Phys. Rev. Appl.}\ }\textbf {\bibinfo {volume} {14}},\
  \bibinfo {pages} {024076} (\bibinfo {year} {2020})}\BibitemShut {NoStop}%
\bibitem [{\citenamefont {Weggler}\ \emph {et~al.}(2020)\citenamefont
  {Weggler}, \citenamefont {Ganslmayer}, \citenamefont {Frank}, \citenamefont
  {Eilert}, \citenamefont {Jelezko},\ and\ \citenamefont
  {Michaelis}}]{Weggler_VecMag2020}%
  \BibitemOpen
  \bibfield  {author} {\bibinfo {author} {\bibfnamefont {T.}~\bibnamefont
  {Weggler}}, \bibinfo {author} {\bibfnamefont {C.}~\bibnamefont {Ganslmayer}},
  \bibinfo {author} {\bibfnamefont {F.}~\bibnamefont {Frank}}, \bibinfo
  {author} {\bibfnamefont {T.}~\bibnamefont {Eilert}}, \bibinfo {author}
  {\bibfnamefont {F.}~\bibnamefont {Jelezko}},\ and\ \bibinfo {author}
  {\bibfnamefont {J.}~\bibnamefont {Michaelis}},\ }\bibfield  {title} {\bibinfo
  {title} {Determination of the three-dimensional magnetic field vector
  orientation with nitrogen vacany centers in diamond},\ }\href
  {https://doi.org/10.1021/acs.nanolett.9b04725} {\bibfield  {journal}
  {\bibinfo  {journal} {Nano Letters}\ }\textbf {\bibinfo {volume} {20}},\
  \bibinfo {pages} {2980} (\bibinfo {year} {2020})}\BibitemShut {NoStop}%
\bibitem [{\citenamefont {Chen}\ \emph {et~al.}(2020)\citenamefont {Chen},
  \citenamefont {Hou}, \citenamefont {Ge}, \citenamefont {Zhang}, \citenamefont
  {Ji}, \citenamefont {Li}, \citenamefont {Qian}, \citenamefont {Wang},
  \citenamefont {Xu},\ and\ \citenamefont {Du}}]{Du_VecMag2020}%
  \BibitemOpen
  \bibfield  {author} {\bibinfo {author} {\bibfnamefont {B.}~\bibnamefont
  {Chen}}, \bibinfo {author} {\bibfnamefont {X.}~\bibnamefont {Hou}}, \bibinfo
  {author} {\bibfnamefont {F.}~\bibnamefont {Ge}}, \bibinfo {author}
  {\bibfnamefont {X.}~\bibnamefont {Zhang}}, \bibinfo {author} {\bibfnamefont
  {Y.}~\bibnamefont {Ji}}, \bibinfo {author} {\bibfnamefont {H.}~\bibnamefont
  {Li}}, \bibinfo {author} {\bibfnamefont {P.}~\bibnamefont {Qian}}, \bibinfo
  {author} {\bibfnamefont {Y.}~\bibnamefont {Wang}}, \bibinfo {author}
  {\bibfnamefont {N.}~\bibnamefont {Xu}},\ and\ \bibinfo {author}
  {\bibfnamefont {J.}~\bibnamefont {Du}},\ }\bibfield  {title} {\bibinfo
  {title} {Calibration-free vector magnetometry using nitrogen-vacancy center
  in diamond integrated with optical vortex beam},\ }\href
  {https://doi.org/10.1021/acs.nanolett.0c03377} {\bibfield  {journal}
  {\bibinfo  {journal} {Nano Letters}\ }\textbf {\bibinfo {volume} {20}},\
  \bibinfo {pages} {8267} (\bibinfo {year} {2020})}\BibitemShut {NoStop}%
\bibitem [{\citenamefont {Reuschel}\ \emph {et~al.}(2022)\citenamefont
  {Reuschel}, \citenamefont {Agio},\ and\ \citenamefont
  {Flatae}}]{Flatae_VecMag}%
  \BibitemOpen
  \bibfield  {author} {\bibinfo {author} {\bibfnamefont {P.}~\bibnamefont
  {Reuschel}}, \bibinfo {author} {\bibfnamefont {M.}~\bibnamefont {Agio}},\
  and\ \bibinfo {author} {\bibfnamefont {A.~M.}\ \bibnamefont {Flatae}},\
  }\bibfield  {title} {\bibinfo {title} {Vector magnetometry based on
  polarimetric optically detected magnetic resonance},\ }\href
  {https://doi.org/https://doi.org/10.1002/qute.202200077} {\bibfield
  {journal} {\bibinfo  {journal} {Advanced Quantum Technologies}\ }\textbf
  {\bibinfo {volume} {5}},\ \bibinfo {pages} {2200077} (\bibinfo {year}
  {2022})}\BibitemShut {NoStop}%
\bibitem [{\citenamefont {Qiu}\ \emph {et~al.}(2021)\citenamefont {Qiu},
  \citenamefont {Vool}, \citenamefont {Hamo},\ and\ \citenamefont
  {Yacoby}}]{Yacoby_AngleSens2021npj}%
  \BibitemOpen
  \bibfield  {author} {\bibinfo {author} {\bibfnamefont {Z.}~\bibnamefont
  {Qiu}}, \bibinfo {author} {\bibfnamefont {U.}~\bibnamefont {Vool}}, \bibinfo
  {author} {\bibfnamefont {A.}~\bibnamefont {Hamo}},\ and\ \bibinfo {author}
  {\bibfnamefont {A.}~\bibnamefont {Yacoby}},\ }\bibfield  {title} {\bibinfo
  {title} {Nuclear spin assisted magnetic field angle sensing},\ }\href
  {https://doi.org/10.1038/s41534-021-00374-6} {\bibfield  {journal} {\bibinfo
  {journal} {npj Quantum Information}\ }\textbf {\bibinfo {volume} {7}},\
  \bibinfo {pages} {39} (\bibinfo {year} {2021})}\BibitemShut {NoStop}%
\bibitem [{\citenamefont {Wang}\ \emph {et~al.}(2015)\citenamefont {Wang},
  \citenamefont {Yuan}, \citenamefont {Huang}, \citenamefont {Rong},
  \citenamefont {Wang}, \citenamefont {Xu}, \citenamefont {Duan}, \citenamefont
  {Ju}, \citenamefont {Shi},\ and\ \citenamefont {Du}}]{DuVectorMW}%
  \BibitemOpen
  \bibfield  {author} {\bibinfo {author} {\bibfnamefont {P.}~\bibnamefont
  {Wang}}, \bibinfo {author} {\bibfnamefont {Z.}~\bibnamefont {Yuan}}, \bibinfo
  {author} {\bibfnamefont {P.}~\bibnamefont {Huang}}, \bibinfo {author}
  {\bibfnamefont {X.}~\bibnamefont {Rong}}, \bibinfo {author} {\bibfnamefont
  {M.}~\bibnamefont {Wang}}, \bibinfo {author} {\bibfnamefont {X.}~\bibnamefont
  {Xu}}, \bibinfo {author} {\bibfnamefont {C.}~\bibnamefont {Duan}}, \bibinfo
  {author} {\bibfnamefont {C.}~\bibnamefont {Ju}}, \bibinfo {author}
  {\bibfnamefont {F.}~\bibnamefont {Shi}},\ and\ \bibinfo {author}
  {\bibfnamefont {J.}~\bibnamefont {Du}},\ }\bibfield  {title} {\bibinfo
  {title} {High-resolution vector microwave magnetometry based on solid-state
  spins in diamond},\ }\href {https://doi.org/10.1038/ncomms7631} {\bibfield
  {journal} {\bibinfo  {journal} {Nature Communications}\ }\textbf {\bibinfo
  {volume} {6}},\ \bibinfo {pages} {6631} (\bibinfo {year} {2015})}\BibitemShut
  {NoStop}%
\bibitem [{\citenamefont {Wang}\ \emph {et~al.}(2021)\citenamefont {Wang},
  \citenamefont {Liu}, \citenamefont {Zhu},\ and\ \citenamefont
  {Cappellaro}}]{Cappellaro_VecAC2021}%
  \BibitemOpen
  \bibfield  {author} {\bibinfo {author} {\bibfnamefont {G.}~\bibnamefont
  {Wang}}, \bibinfo {author} {\bibfnamefont {Y.-X.}\ \bibnamefont {Liu}},
  \bibinfo {author} {\bibfnamefont {Y.}~\bibnamefont {Zhu}},\ and\ \bibinfo
  {author} {\bibfnamefont {P.}~\bibnamefont {Cappellaro}},\ }\bibfield  {title}
  {\bibinfo {title} {Nanoscale vector ac magnetometry with a single
  nitrogen-vacancy center in diamond},\ }\href
  {https://doi.org/10.1021/acs.nanolett.1c01165} {\bibfield  {journal}
  {\bibinfo  {journal} {Nano Letters}\ }\textbf {\bibinfo {volume} {21}},\
  \bibinfo {pages} {5143} (\bibinfo {year} {2021})}\BibitemShut {NoStop}%
\bibitem [{\citenamefont {Zhang}\ and\ \citenamefont
  {Suter}(2023)}]{Jingfu_RFsens2023}%
  \BibitemOpen
  \bibfield  {author} {\bibinfo {author} {\bibfnamefont {J.}~\bibnamefont
  {Zhang}}\ and\ \bibinfo {author} {\bibfnamefont {D.}~\bibnamefont {Suter}},\
  }\bibfield  {title} {\bibinfo {title} {Single nv centers as sensors for
  radio-frequency fields},\ }\href
  {https://doi.org/10.1103/PhysRevResearch.5.L022026} {\bibfield  {journal}
  {\bibinfo  {journal} {Phys. Rev. Res.}\ }\textbf {\bibinfo {volume} {5}},\
  \bibinfo {pages} {L022026} (\bibinfo {year} {2023})}\BibitemShut {NoStop}%
\bibitem [{\citenamefont {Lamba}\ \emph {et~al.}(2024)\citenamefont {Lamba},
  \citenamefont {Rana}, \citenamefont {Halder}, \citenamefont {Dhomkar},
  \citenamefont {Suter},\ and\ \citenamefont {Kamineni}}]{PoojaVecSens2024}%
  \BibitemOpen
  \bibfield  {author} {\bibinfo {author} {\bibfnamefont {P.}~\bibnamefont
  {Lamba}}, \bibinfo {author} {\bibfnamefont {A.}~\bibnamefont {Rana}},
  \bibinfo {author} {\bibfnamefont {S.}~\bibnamefont {Halder}}, \bibinfo
  {author} {\bibfnamefont {S.}~\bibnamefont {Dhomkar}}, \bibinfo {author}
  {\bibfnamefont {D.}~\bibnamefont {Suter}},\ and\ \bibinfo {author}
  {\bibfnamefont {R.~K.}\ \bibnamefont {Kamineni}},\ }\bibfield  {title}
  {\bibinfo {title} {Vector detection of ac magnetic fields by nitrogen vacancy
  centers of single orientation in diamond},\ }\href
  {https://doi.org/10.1103/PhysRevB.109.195424} {\bibfield  {journal} {\bibinfo
   {journal} {Phys. Rev. B}\ }\textbf {\bibinfo {volume} {109}},\ \bibinfo
  {pages} {195424} (\bibinfo {year} {2024})}\BibitemShut {NoStop}%
\bibitem [{\citenamefont {Cai}\ \emph {et~al.}(2024)\citenamefont {Cai},
  \citenamefont {Weng}, \citenamefont {Zhu}, \citenamefont {Zhu}, \citenamefont
  {Lou},\ and\ \citenamefont {Wang}}]{WangVecMWmag2024}%
  \BibitemOpen
  \bibfield  {author} {\bibinfo {author} {\bibfnamefont {M.}~\bibnamefont
  {Cai}}, \bibinfo {author} {\bibfnamefont {C.}~\bibnamefont {Weng}}, \bibinfo
  {author} {\bibfnamefont {Y.}~\bibnamefont {Zhu}}, \bibinfo {author}
  {\bibfnamefont {W.}~\bibnamefont {Zhu}}, \bibinfo {author} {\bibfnamefont
  {L.}~\bibnamefont {Lou}},\ and\ \bibinfo {author} {\bibfnamefont
  {G.}~\bibnamefont {Wang}},\ }\bibfield  {title} {\bibinfo {title} {Using a
  single nitrogen-vacancy center in diamond to detect microwave magnetic field
  vectors at resonant frequency},\ }\href {https://doi.org/10.1063/5.0197740}
  {\bibfield  {journal} {\bibinfo  {journal} {Applied Physics Letters}\
  }\textbf {\bibinfo {volume} {124}},\ \bibinfo {pages} {154001} (\bibinfo
  {year} {2024})}\BibitemShut {NoStop}%
\bibitem [{\citenamefont {Moreva}\ \emph {et~al.}(2020)\citenamefont {Moreva},
  \citenamefont {Bernardi}, \citenamefont {Traina}, \citenamefont {Sosso},
  \citenamefont {Tchernij}, \citenamefont {Forneris}, \citenamefont {Picollo},
  \citenamefont {Brida}, \citenamefont {Pastuovi\ifmmode~\acute{c}\else
  \'{c}\fi{}}, \citenamefont {Degiovanni}, \citenamefont {Olivero},\ and\
  \citenamefont {Genovese}}]{Genovese_TempSens}%
  \BibitemOpen
  \bibfield  {author} {\bibinfo {author} {\bibfnamefont {E.}~\bibnamefont
  {Moreva}}, \bibinfo {author} {\bibfnamefont {E.}~\bibnamefont {Bernardi}},
  \bibinfo {author} {\bibfnamefont {P.}~\bibnamefont {Traina}}, \bibinfo
  {author} {\bibfnamefont {A.}~\bibnamefont {Sosso}}, \bibinfo {author}
  {\bibfnamefont {S.~D.}\ \bibnamefont {Tchernij}}, \bibinfo {author}
  {\bibfnamefont {J.}~\bibnamefont {Forneris}}, \bibinfo {author}
  {\bibfnamefont {F.}~\bibnamefont {Picollo}}, \bibinfo {author} {\bibfnamefont
  {G.}~\bibnamefont {Brida}}, \bibinfo {author} {\bibfnamefont {i.~c.~v.}\
  \bibnamefont {Pastuovi\ifmmode~\acute{c}\else \'{c}\fi{}}}, \bibinfo {author}
  {\bibfnamefont {I.~P.}\ \bibnamefont {Degiovanni}}, \bibinfo {author}
  {\bibfnamefont {P.}~\bibnamefont {Olivero}},\ and\ \bibinfo {author}
  {\bibfnamefont {M.}~\bibnamefont {Genovese}},\ }\bibfield  {title} {\bibinfo
  {title} {Practical applications of quantum sensing: A simple method to
  enhance the sensitivity of nitrogen-vacancy-based temperature sensors},\
  }\href {https://doi.org/10.1103/PhysRevApplied.13.054057} {\bibfield
  {journal} {\bibinfo  {journal} {Phys. Rev. Appl.}\ }\textbf {\bibinfo
  {volume} {13}},\ \bibinfo {pages} {054057} (\bibinfo {year}
  {2020})}\BibitemShut {NoStop}%
\bibitem [{\citenamefont {Bitter}(1949)}]{Bitter_ODMR1949}%
  \BibitemOpen
  \bibfield  {author} {\bibinfo {author} {\bibfnamefont {F.}~\bibnamefont
  {Bitter}},\ }\bibfield  {title} {\bibinfo {title} {The optical detection of
  radiofrequency resonance},\ }\href {https://doi.org/10.1103/PhysRev.76.833}
  {\bibfield  {journal} {\bibinfo  {journal} {Phys. Rev.}\ }\textbf {\bibinfo
  {volume} {76}},\ \bibinfo {pages} {833} (\bibinfo {year} {1949})}\BibitemShut
  {NoStop}%
\bibitem [{\citenamefont {Suter}(2020)}]{Suter_ODMRRev2020}%
  \BibitemOpen
  \bibfield  {author} {\bibinfo {author} {\bibfnamefont {D.}~\bibnamefont
  {Suter}},\ }\bibfield  {title} {\bibinfo {title} {Optical detection of
  magnetic resonance},\ }\href {https://doi.org/10.5194/mr-1-115-2020}
  {\bibfield  {journal} {\bibinfo  {journal} {Magnetic Resonance}\ }\textbf
  {\bibinfo {volume} {1}},\ \bibinfo {pages} {115} (\bibinfo {year}
  {2020})}\BibitemShut {NoStop}%
\bibitem [{\citenamefont {Shim}\ \emph {et~al.}(2013)\citenamefont {Shim},
  \citenamefont {Nowak}, \citenamefont {Niemeyer}, \citenamefont {Zhang},
  \citenamefont {Brandao},\ and\ \citenamefont {~}}]{ShimArXiv}%
  \BibitemOpen
  \bibfield  {author} {\bibinfo {author} {\bibfnamefont {J.~H.}\ \bibnamefont
  {Shim}}, \bibinfo {author} {\bibfnamefont {B.}~\bibnamefont {Nowak}},
  \bibinfo {author} {\bibfnamefont {I.}~\bibnamefont {Niemeyer}}, \bibinfo
  {author} {\bibfnamefont {J.}~\bibnamefont {Zhang}}, \bibinfo {author}
  {\bibfnamefont {F.~D.}\ \bibnamefont {Brandao}},\ and\ \bibinfo {author}
  {\bibfnamefont {D.}~\bibnamefont {~}},\ }\href
  {http://arxiv.org/abs/1307.0257} {\bibfield  {journal} {\bibinfo  {journal}
  {arXiv:1307.0257 [quant-ph]}\ } (\bibinfo {year} {2013})}\BibitemShut
  {NoStop}%
\bibitem [{\citenamefont {Maletinsky}\ \emph {et~al.}(2012)\citenamefont
  {Maletinsky}, \citenamefont {Hong}, \citenamefont {Grinolds}, \citenamefont
  {Hausmann}, \citenamefont {Lukin}, \citenamefont {Walsworth}, \citenamefont
  {Loncar},\ and\ \citenamefont {Yacoby}}]{Maletinsky_NVAFM2012}%
  \BibitemOpen
  \bibfield  {author} {\bibinfo {author} {\bibfnamefont {P.}~\bibnamefont
  {Maletinsky}}, \bibinfo {author} {\bibfnamefont {S.}~\bibnamefont {Hong}},
  \bibinfo {author} {\bibfnamefont {M.~S.}\ \bibnamefont {Grinolds}}, \bibinfo
  {author} {\bibfnamefont {B.}~\bibnamefont {Hausmann}}, \bibinfo {author}
  {\bibfnamefont {M.~D.}\ \bibnamefont {Lukin}}, \bibinfo {author}
  {\bibfnamefont {R.~L.}\ \bibnamefont {Walsworth}}, \bibinfo {author}
  {\bibfnamefont {M.}~\bibnamefont {Loncar}},\ and\ \bibinfo {author}
  {\bibfnamefont {A.}~\bibnamefont {Yacoby}},\ }\bibfield  {title} {\bibinfo
  {title} {A robust scanning diamond sensor for nanoscale imaging with single
  nitrogen-vacancy centres},\ }\href {https://doi.org/10.1038/nnano.2012.50}
  {\bibfield  {journal} {\bibinfo  {journal} {Nature Nanotechnology}\ }\textbf
  {\bibinfo {volume} {7}},\ \bibinfo {pages} {320} (\bibinfo {year}
  {2012})}\BibitemShut {NoStop}%
\end{thebibliography}%

\end{document}